%% file: MiNNLO_nunugam.tex
\documentclass[12pt,a4paper]{article}
\pdfoutput=1
\usepackage[utf8]{inputenc}
\newcommand{\sphid}[1]{}

\usepackage{epsf,amsmath,amssymb,graphicx,dcolumn}
\usepackage{caption}
\usepackage[labelformat=simple]{subcaption}

\usepackage{scalefnt,ulem,pstricks}
\usepackage{booktabs,multirow,tabularx}
\usepackage[colorlinks=true,allcolors={blue!70!black}]{hyperref}
\usepackage{cleveref}
\usepackage{color}
\usepackage{rotating}
\usepackage{microtype}
\usepackage[titletoc,title]{appendix}
\usepackage[numbers,sort&compress]{natbib}
\usepackage{amsmath,amsfonts,amsthm,bm}
\usepackage{xspace}
\input{definitions.tex}

\usepackage[tikz]{bclogo}
\usepackage{subcaption}
\usepackage{tikz-feynman}
\usepackage{cancel}
\usepackage{verbatim}
\usetikzlibrary{arrows,shapes}
\usepackage{afterpage}

\begin{document} 
\begin{flushright}
\vspace*{-1.5cm}
MPP-2021-138
\end{flushright}
\vspace{0.cm}

\begin{center}
{\Large \bf Anomalous couplings in \boldmath{$Z\gamma$} events at NNLO+PS\\[0.2cm] and improving $\nu\bar\nu\gamma$ backgrounds in dark-matter searches}
\end{center}

\begin{center}
{\bf Daniele Lombardi}, {\bf Marius Wiesemann}, and {\bf Giulia Zanderighi}

Max-Planck-Institut f\"ur Physik, F\"ohringer Ring 6, 80805 M\"unchen, Germany

\href{mailto:lombardi@mpp.mpg.de}{\tt lombardi@mpp.mpg.de}\\
\href{mailto:marius.wiesemann@mpp.mpg.de}{\tt marius.wiesemann@mpp.mpg.de}\\
\href{mailto:zanderi@mpp.mpg.de}{\tt zanderi@mpp.mpg.de}

\end{center}

\begin{center} {\bf Abstract} \end{center}\vspace{-1cm}
\begin{quote}
\pretolerance 10000
The measurement of the triple gauge couplings (TGCs) is a central 
part of diboson studies at the LHC. In this letter we consider 
the $Z\gamma$ process and include anomalous TGCs (aTGCs)
in the event generation at next-to-next-to-leading order QCD accuracy (NNLO+PS) 
within the \minnlo{} framework. While our implementation 
is fully general and applies to both $Z\to \ell^+\ell^-$
and $Z\to \nu\bar\nu$ decays, we focus here on the $\nu\bar\nu\gamma$ final state.
After validation of our simulation of $\nu\bar\nu\gamma$ events, for which 
NNLO+PS accuracy is achieved for the first time, the effects 
of aTGCs on various distributions are studied. Moreover, we show 
the relevance of NNLO+PS accuracy for the $\nu\bar\nu\gamma$ 
background to photon plus missing energy signatures in dark-matter searches,
and we compare \minnlo{} predictions for $\nu\bar\nu\gamma$ production to recent $13$\,TeV data.
\end{quote}

\parskip = 1.2ex

Precision phenomenology has evolved to one of the cornerstones
of the rich physics programme at the Large Hadron Collider (LHC).
Without a clear signal of new-physics phenomena, the accurate measurement 
of rates and distributions of Standard Model (SM) processes provides a 
potential pathway towards the discovery of physics beyond the SM (BSM) 
through small deviations from the SM predictions. Vector-boson pair production
processes represent an important family of reactions in that respect, as they provide 
direct access to anomalous couplings between three gauge bosons 
(anomalous triple gauge coupling, or aTGC). Constraining or finding
anomalous couplings profits directly from high-accuracy computations in 
perturbation theory.
Indeed an enormous effort in the recent years 
has been spent on next-to-next-to-leading order (NNLO) 
QCD calculations, which are the standard today for 
colour-singlet production involving up to two bosons~\cite{Ferrera:2011bk,Ferrera:2014lca,Ferrera:2017zex,Campbell:2016jau,Harlander:2003ai,Harlander:2010cz,Harlander:2011fx,Buehler:2012cu,Marzani:2008az,Harlander:2009mq,Harlander:2009my,Pak:2009dg,Neumann:2014nha,Catani:2011qz,Campbell:2016yrh,Grazzini:2013bna,Grazzini:2015nwa,Campbell:2017aul,Gehrmann:2020oec,Cascioli:2014yka,Grazzini:2015hta,Heinrich:2017bvg,Kallweit:2018nyv,Gehrmann:2014fva,Grazzini:2016ctr,Grazzini:2016swo,Grazzini:2017ckn,deFlorian:2013jea,deFlorian:2016uhr,Grazzini:2018bsd,Baglio:2012np,Li:2016nrr,deFlorian:2019app}. Within the last two years even the first $2\to 3$ LHC processes have been 
computed at NNLO QCD, namely $\gamma\gamma\gamma$~\cite{Chawdhry:2019bji,Kallweit:2020gcp}
$\gamma\gamma$+jet~\cite{Czakon:2021mjy} and
3-jet~\cite{Chawdhry:2021hkp} production. Indeed,
in almost all cases NNLO corrections turn out to be crucial 
to accurately describe data from LHC measurements 
within the experimental uncertainties.

The production of a $Z$ boson and a photon ($Z\gamma$ production) 
is an important vector-boson pair production process in various respects.
First of all, the measurement of non-zero $ZZ\gamma$ or $Z\gamma\gamma$ couplings, which are absent in the
SM, would be direct evidence of BSM physics. 
Moreover, $Z\gamma$ final states are relevant in direct searches for BSM
particles. In particular, in the $Z\to\nu\bar\nu$ decay channel $Z\gamma$ production
constitutes an irreducible background to dark-matter searches in the photon plus missing 
energy final state. The low accuracy of available  $\nu\bar\nu\gamma$ event simulations
 is actually one of the limiting factors in current dark-matter analyses \cite{ATLAS:2020uiq}, despite the 
 fact that a substantial effort has been made to improve the perturbative accuracy of 
 $Z\gamma$ production. Next-to-leading order (NLO)
QCD corrections have been known for a long time for on-shell $Z$ bosons \cite{Ohnemus:1992jn} 
and including their leptonic decays \cite{Baur:1997kz,Campbell:2011bn}. 
Also the NNLO QCD cross section has been calculated at the fully differential level and including leptonic decays \cite{Grazzini:2013bna,Grazzini:2015nwa,Campbell:2017aul} as well as 
 NLO electroweak (EW) corrections \cite{Hollik:2004tm,Accomando:2005ra}.
More recently, the resummation of large logarithmic corrections for this process has been 
combined with NNLO QCD predictions \cite{Kallweit:2020gva,Wiesemann:2020gbm,Becher:2020ugp}.

In this letter we consider $Z\gamma$ production 
in the $Z\to\nu\bar\nu$ decay channel, and we 
present NNLO QCD predictions matched to parton showers (NNLO+PS)
for the $\nu\bar\nu\gamma$ final state.
This calculation extends the list of LHC production processes 
available at NNLO+PS, which includes Higgs-boson~\cite{Hamilton:2013fea,Hoche:2014dla,Monni:2019whf,Monni:2020nks}, Drell-Yan~\cite{Hoeche:2014aia,Karlberg:2014qua,Alioli:2015toa,Monni:2019whf,Monni:2020nks}, Higgsstrahlung~\cite{Astill:2016hpa,Astill:2018ivh,Alioli:2019qzz}, 
$\gamma\gamma$~\cite{Alioli:2020qrd}, $\ell^+\ell^-\gamma$~\cite{Lombardi:2020wju}, $\ell^\pm \nu \gamma$~\cite{Cridge:2021hfr}, $\ell^+\nu_\ell \ell^{\prime-}\nu_{\ell^{\prime-}}$~\cite{Re:2018vac}, $\ell^+\ell^-\ell^{\prime+}\ell^{\prime-}$~\cite{Buonocore:2021fnj,Alioli:2021egp} and $t\bar{t}$ production~\cite{Mazzitelli:2020jio} so far.
These simulations rely on overall four different NNLO+PS approaches that 
were developed in the last years
\cite{Hamilton:2012rf,Alioli:2013hqa,Hoeche:2014aia,Monni:2019whf,Monni:2020nks, Mazzitelli:2020jio}. Our calculation employs the \minnlo{} approach of \citeres{Monni:2019whf,Monni:2020nks} and it is based on the 
$Z\gamma$ \minnlo{} generator for $\ell^+\ell^-\gamma$ production 
presented in \citere{Lombardi:2020wju}. To the purpose of our study, 
we have extended the previous implementation to
deal with the $\nu\bar\nu\gamma$ final state 
and we have included in the event generation the effects of aTGCs, specifically 
the $ZZ\gamma$ and $Z\gamma\gamma$ verteces.\footnote{Although 
we study the $\nu\bar\nu\gamma$ process here, the implementation of the 
aTGCs can readily be used to include those effects in the $\ell^+\ell^-\gamma$  event generation at NNLO+PS as well.}

We recall that \minnlo{} is a powerful approach with various positive features:
NNLO corrections are calculated directly during event generation, 
without the need for an a-posteriori reweighting. Moreover, no
merging scale or slicing cutoff is required to separate different 
multiplicities in the generated event samples, 
keeping power-suppressed terms into account.
The leading-logarithmic accuracy of the shower is preserved when
combined with transverse-momentum ordered parton showers.
Although the \minnlo{} method has been initially developed on 
basis of the transverse momentum of colour singlet, the idea behind 
the approach is neither limited to a specific observable nor to colour-singlet 
production. Indeed, the method has recently been extended to heavy-quark
pair production in \citere{Mazzitelli:2020jio}.

\begin{figure}[t]
  \begin{center}
    \begin{subfigure}[b]{.5\linewidth}
      \centering
\begin{tikzpicture}
\begin{feynman}
        \vertex (a1) {\( q\)};
        \vertex[below=1.6cm of a1] (a2){\(\overline q\)};
        \vertex[right=2cm of a1] (a3);
        \vertex[right=2cm of a2] (a4);
        \vertex[right=1.65cm of a3] (a5){\(\gamma\)};
        \vertex[right=2cm of a3] (a9);
        \vertex[right=1cm of a4] (a6);
        \vertex[below=0.7cm of a9] (a7){\(\nu\)} ;
        \vertex[below=1.7cm of a9] (a8){\(\overline\nu\)};
        
        \diagram* {
          {[edges=fermion]
            (a1)--(a3)--(a4)--(a2),
            (a7)--(a6)--(a8),
          },
          (a3) -- [ boson] (a5),
          (a4) -- [boson, edge label'=\(Z\)] (a6),
        };

      \end{feynman}
\end{tikzpicture}
\caption{$q\bar{q}$-initiated $t$-channel diagram}
        \label{subfig:qq}
\end{subfigure}%
\begin{subfigure}[b]{.5\linewidth}
  \centering
\begin{tikzpicture}

  \begin{feynman}
    \vertex (a1) {\( g\)};
    \vertex[below=1.6cm of a1] (a2){\(g\)};
    \vertex[right=1.5cm of a1] (a3);
    \vertex[right=1.5cm of a2] (a4);
    \vertex[right=1cm of a3] (a5);
    \vertex[right=1cm of a4] (a6);
    \vertex[right=1.5cm of a5] (a7){\(\gamma\)};
    \vertex[right=1.9cm of a5] (a11);
    \vertex[right=0.9cm of a6] (a8);
    \vertex[below=0.7cm of a11] (a9){\(\nu\)} ;
    \vertex[below=1.7cm of a11] (a10){\(\bar\nu\)};
 
    \diagram* {
      {[edges=fermion]
        (a3)--(a4)--(a6)--(a5)--(a3),
        (a9)--(a8)--(a10),
      },
      (a1) -- [ gluon] (a3),
      (a2) -- [ gluon] (a4),
      (a5) -- [ boson] (a7),
      (a6) -- [boson, edge label'=\(Z\)] (a8),
       };

  \end{feynman}
\end{tikzpicture}
\caption{$gg$-initiated loop-induced diagram}
        \label{subfig:gg}
\end{subfigure}
\end{center}
\caption{\label{DiagramsZgamma} Sample Feynman diagrams for 
  $\nu\bar\nu\gamma$ production entering at (a) LO and (b) NNLO.}
\end{figure}
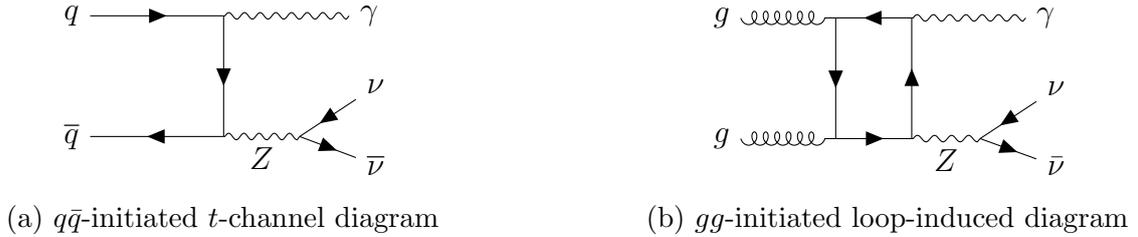

We consider the process
\begin{align}
\label{eq:proc}
pp\to\nu\bar\nu\gamma + X\,,
\end{align}
with $\nu\in\{\nu_e,\nu_\mu,\nu_\tau\}$
and by accounting for all relevant topologies leading to this final state in our calculation, 
we include interferences, off-shell effects and spin correlations. At leading order (LO), the process is quark--anti-quark ($q\bar{q}$) induced in the SM
and proceeds only via a $t$-channel quark exchange with both the isolated photon and the $Z$ boson coupling to the quark line. 
This is different from the $\ell^+\ell^-\gamma$ final state, where Drell-Yan-like $s$-channel topologies are allowed as well, 
since the charged leptons can emit an isolated photon, while the
neutrinos can not. A representative LO Feynman diagram is shown in \fig{subfig:qq}.
At NNLO in QCD perturbation theory the loop-induced 
gluon-fusion ($gg$) contribution enters the cross section, see \fig{subfig:gg}. 
However, this contribution is very 
small -- at the (sub-)percent level -- and will be neglected throughout this letter, as it 
can be calculated completely independently from the $q\bar{q}$ initiated process.

We calculate NNLO+PS predictions for the process in \eqn{eq:proc} by means of the 
\minnlo{} method. Our implementation is based on the $Z\gamma$ \minnlo{} generator
developed in \citere{Lombardi:2020wju}, which has been extended to the $\nu\bar\nu\gamma$ final state.
As for the $\ell^+\ell^-\gamma$ final state discussed in \citere{Lombardi:2020wju}, tree-level and one-loop
amplitudes can be evaluated either through an analytic implementation taken from \noun{MCFM}~\cite{Campbell:2019dru} or through \OpenLoops{}~\cite{Cascioli:2011va,Buccioni:2017yxi,Buccioni:2019sur}, while for the two-loop amplitude we 
rely on its implementation within the \Matrix{} framework~\cite{Grazzini:2017mhc,Matrixurl} that is based on the calculation of \citere{Gehrmann:2011ab}.\footnote{Note that we have corrected a minor sign mistake in the analytic continuation of a complex logarithm entering the two-loop quark form factor in \Matrix{}.}
The \minnlo{} method has been formulated in \citere{Monni:2019whf}, optimised for $2\rightarrow1$ processes in \citere{Monni:2020nks} and later extended to generic colour-singlet processes in \citere{Lombardi:2020wju} and to heavy-quark pair production in \citere{Mazzitelli:2020jio}. We refer to those publications for details. The \minnlo{} master formula can be symbolically 
expressed through the \POWHEG{} \cite{Nason:2004rx,Nason:2006hfa,Frixione:2007vw,Alioli:2010xd} formula for the production of a system of colour singlets (\F{}) plus one light parton 
(\FJ{}):
\begin{align}
\label{eq:master}
{\rm d}\sigma_{\rm\scriptscriptstyle F}^{\rm MiNNLO_{PS}}=\bar{B}^{\,\rm MiNNLO_{\rm PS}}\,\times\,\left\{\Delta_{\rm pwg}(\Lambda_{\rm pwg})+\int {\rm d}\Phi_{\rm rad}\Delta_{\rm pwg}(p_{T,{\rm rad}})\,\frac{R_{\scriptscriptstyle\rm FJ}}{B_{\scriptscriptstyle\rm FJ}}\right\}\,,
\end{align}
with a modified content of the $\bar B$ function
\begin{align}
\label{eq:minnlo}
\bar{B}^{\,\rm MiNNLO_{\rm PS}}\sim e^{-S}\,\Big\{{\rm d}\sigma^{(1)}_{\scriptscriptstyle\rm FJ}\big(1+S^{(1)}\big)+{\rm d}\sigma^{(2)}_{\scriptscriptstyle\rm FJ}+\left(D-D^{(1)}-D^{(2)}\right)\times F^{\rm corr}\Big\}\,,
\end{align}
which ensures that that NNLO accuracy for $\F{}$ production is achieved when the additional jet becomes unresolved.
In \eqn{eq:master} we denote with $\Delta_{\rm pwg}$ the \POWHEG{} Sudakov form factor, and with $\Phi_{\tmop{rad}} $ and $\ptrad$ 
the phase space and the transverse momentum of the second radiation. $B_{\scriptscriptstyle\rm FJ}$ and $R_{\scriptscriptstyle\rm FJ}$ are the squared tree-level matrix elements for \FJ{} and \FJJ{} production, respectively.
${\rm d}\sigma^{(1,2)}_{\scriptscriptstyle\rm FJ}$ in \eqn{eq:minnlo} are the first- and second-order contribution 
to the differential \FJ{} cross section and $e^{-S}$ denotes the Sudakov form factor for the transverse momentum ($\pt$) of the 
colour singlet. The renormalization and factorization scales are evaluated as $\muR\sim\muF\sim \pt$ in \minnlo{}.
NNLO accuracy is achieved through the third term in \eqn{eq:minnlo}, which adds the relevant (singular) contributions of order $\as^3(p_{\text{\scalefont{0.77}T}})$~\cite{Monni:2019whf}. Regular contributions at this order are of subleading nature.
Those contributions are derived from the (fully differential) transverse-momentum resummation formula that can be written (approximately in direct space) as
\begin{align}
\label{eq:resum}
{\rm d}\sigma_{\scriptscriptstyle\rm F}^{\rm res}=\frac{{\rm d}}{{\rm d}\pt}\left\{e^{-S}\mathcal{L}\right\}=e^{-S}\underbrace{\left\{-S^\prime\mathcal{L}+\mathcal{L}^\prime\right\}}_{\equiv D}\,,
\end{align}
which defines the function $D$ in \eqn{eq:minnlo}. Here, $\mathcal{L}$ is the luminosity factor up to NNLO including 
the convolution of the collinear coefficient functions with the parton distribution functions (PDFs) and the squared hard-virtual matrix elements for \F{} production up to two-loop order.
In fact, \eqn{eq:minnlo} follows directly from the matching of \eqn{eq:resum} to the fixed-order cross section ${\rm d}\sigma_{\scriptscriptstyle\rm FJ}$ in a matching scheme where the Sudakov form factor is factored out.
Finally, $F^{\rm corr}$ in \eqn{eq:minnlo} represents the appropriate function to spread the NNLO corrections in the 
\FJ{} phase space, which is necessary to include those corrections in the context of the \FJ{} \POWHEG{} calculation. 

In summary, the \minnlo{} procedure involves essentially three steps: first, the \FJ{} final state is described
at NLO accuracy using \POWHEG{}, inclusively over the radiation of a second light parton. Second, the limit in which 
the light partons become unresolved is corrected by supplementing the appropriate Sudakov form factor and 
higher-order terms, such that the simulation remains finite as well as NNLO accurate for inclusive
\F{} production. These first two steps are included in the $\bar{B}$ function of \eqn{eq:minnlo}.
Third, the second radiation is generated exclusively through the
content of the curly brackets in \eqn{eq:master}, with a default cutoff of $\LambdaPWG=0.89$\,GeV,
preserving the NLO accuracy of \FJ{} production, and subsequent radiation is included through 
the parton shower. We stress that \minnlo{} preserves the (leading logarithmic) 
accuracy of the parton shower, since the analytic Sudakov matches 
the leading logarithms generated by the parton shower itself and since all emissions are appropriately
ordered (when matching to a $p_T$-ordered shower).

The ensuing calculation allows us to retain NNLO QCD
accuracy in the event generation for $\nu\bar\nu\gamma$ production 
interfaced to a parton shower, which is necessary for a complete and realistic event
simulation.  In particular, multiple photon emissions through a QED
shower, as well as non-perturbative QCD effects using hadronization and
underlying event models can be included.  It is well known that these corrections can have a
substantial impact on the lepton momenta, jet-binned cross sections and other more
exclusive observables measured at the LHC.

In addition to the SM simulation, we have implemented the leading 
contributions from aTGCs in the $Z\gamma$ \minnlo{} generator for both 
the $\ell^+\ell^-\gamma$ and the $\nu\bar\nu\gamma$ final states, but we will 
focus on the latter when presenting phenomenological results below.
Since couplings between three charge-neutral weak bosons are forbidden by the SM gauge symmetry, 
their contributions can arise only from BSM theories. This is one
of the reasons why they provide a powerful way of searching for new physics.
Extensions of the standard model gauge structure through aTGC can be described by means of
two equivalent approaches~\cite{Baur:1992cd, DeFlorian:2000sg, Hagiwara:1986vm, Gounaris:1999kf}: the vertex-function and the Lagrangian approach. Both descriptions
can be embedded in a self-consistent effective-field-theory (EFT) framework, as presented in
\citeres{Degrande:2012wf, Degrande:2013kka}. Here, we follow the vertex-function approach, 
as this is usually employed by the experimental analyses, see for instance \citere{ATLAS:2018nci}.

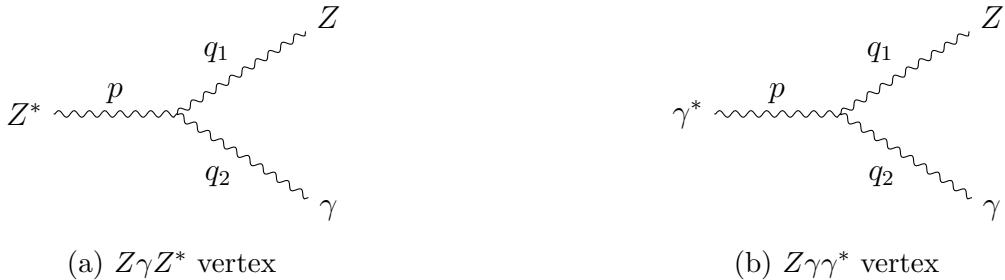
\begin{figure}[t]
  \begin{center}
    \begin{subfigure}[b]{.5\linewidth}
      \centering
\begin{tikzpicture}
  \begin{feynman}
    \vertex (a1) {\( Z^*\)};
    \vertex[right=2cm of a1] (a2);
    \vertex[right=2cm of a2] (a3);
    \vertex[above=1cm of a3] (a4){\(Z\)} ;
    \vertex[below=1cm of a3] (a5){\(\gamma\)};
 
    \diagram* {
      (a1) -- [ boson, edge label = \(p\)] (a2),
      (a2) -- [ boson, edge label = \(q_1\)] (a4),
      (a2) -- [ boson, edge label' = \(q_2\)] (a5),
       };

  \end{feynman}
\end{tikzpicture}
\caption{$Z\gamma Z^*$ vertex}
        \label{subfig:ZgZ}
\end{subfigure}%
\begin{subfigure}[b]{.5\linewidth}
  \centering
\begin{tikzpicture}
  \begin{feynman}
    \vertex (a1) {\( \gamma^*\)};
    \vertex[right=2cm of a1] (a2);
    \vertex[right=2cm of a2] (a3);
    \vertex[above=1cm of a3] (a4){\(Z\)} ;
    \vertex[below=1cm of a3] (a5){\(\gamma\)};
 
    \diagram* {
      (a1) -- [ boson, edge label = \(p\)] (a2),
      (a2) -- [ boson, edge label = \(q_1\)] (a4),
      (a2) -- [ boson, edge label' = \(q_2\)] (a5),
       };

  \end{feynman}

\end{tikzpicture}
\caption{$Z\gamma\gamma^*$ vertex}
        \label{subfig:Zgg}
\end{subfigure}
\end{center}
  \caption{\label{fig:anomvertex} Anomalous couplings between three gauge bosons that are relevant for
  $Z\gamma$ production.}
\end{figure}

For $Z\gamma$ production, two different neutral aTGCs enter the cross section, namely $Z\gamma V$ with $V=Z^*,\,\gamma^*$, which are 
shown in Fig~\ref{fig:anomvertex}. The form of these effective interactions can be constraint
by imposing Lorentz and electromagnetic gauge invariance, as well as Bose statistics.
The latter for instance forbids vertices such as $Z\gamma Z^*$ or $Z\gamma\gamma^*$ with all
gauge bosons being on-shell, since in either case two identical particles are involved in the interaction.
Moreover, we include only terms of dimension less than or equal to eight for practical reasons, as discussed 
in \citere{DeFlorian:2000sg}. This choice, which avoids the proliferation
of couplings that are in principle allowed by symmetry, is 
justified from an EFT perspective, where only a limited amount of
higher-dimensional operators is expected to contribute to the physical process at a given
energy scale.
With these minimal requirements, the effective interaction can be parametrized as \cite{Hagiwara:1986vm}:
\begin{align}
  \label{eq:effvertex}
\Gamma^{\alpha \beta \mu}_{Z \gamma V}(q_1, q_2, p) &=  
\frac{i(p^2-m_V^2)}{\Lambda^2} \Biggl( 
h_1^V \bigl( q_2^\mu g^{\alpha\beta} - q_2^\alpha g^{\mu \beta}
\bigr) + \notag\\
 & +\frac{h_2^V}{\Lambda^2} p^\alpha \Bigl( p\cdot q_2\ g^{\mu\beta} -
q_2^\mu p^\beta \Bigr)
- h_3^V \varepsilon^{\mu\alpha\beta\nu} q_{2\, \nu} 
- \frac{h_4^V}{\Lambda^2} \varepsilon^{\mu\beta\nu\sigma} p^\alpha
p_\nu q_{2\, \sigma} \Biggl)\,,
\end{align}
where $q_1$ and $q_2$ are the momenta of the on-shell $Z$ and $\gamma$ gauge bosons,
respectively, and $p$ is the momentum of the off-shell boson $V$. One should bear in mind that 
in principle additional terms arise when all gauge bosons are considered to be off shell. However,
since the $Z\gamma$ analyses select isolated photons and measure predominantly contributions 
from $Z$ bosons close to their mass shell, \eqn{eq:effvertex} provides the dominant effects 
also when including the leptonic decay of the final-state $Z$ boson.
The two anomalous vertices $Z\gamma Z^*$ and $Z\gamma\gamma^*$ are obtained by choosing the $V=Z^*,\,\gamma^*$
and setting $m^2_V$ accordingly. The effective couplings parametrizing the interaction
are given by $h_i^V$ with $i\in\{1\dots4\}$ in \eqn{eq:effvertex} and $\Lambda$ is a mass scale
conventionally chosen to be the $Z$ boson mass $m_Z$. Note that a different scale choice for
$\Lambda$ just amounts to a rescaling of all $h_i^V$ couplings~\cite{Baur:1992cd}.

The $h_1^V$ and $h_2^V$ anomalous couplings are CP violating, and would only
appear in a UV completion of the SM allowing for new particles with CP violating
interactions with the SM ones. Being CP odd, these terms can not interfere with the SM
sector and they can just contribute to the cross section at quadratic level. 
On the contrary,
the CP-preserving couplings $h_3^V$ and $h_4^V$ enter the cross section also with linear
terms through interference with the SM amplitudes.
In principle one may think that this fact renders the CP-violating couplings more 
difficult to constrain \cite{Gounaris:1999kf}; however, the linear term involves an interference between the $t$-channel and $s$-channel diagrams (of the SM and BSM contribution, respectively), which is strongly suppressed.
Note also that the experimental sensitivity is affected by the dimensionality of the coupling itself. 
In particular, $h_{2/4}^V$ induce dimension eight terms, which grow with two extra powers of the energy scale 
with respect to the dimension six couplings $h_{1/3}^V$. Thus better limits can be obtained for $h_{2/4}^V$ \cite{Gounaris:1999kf}. 
There are many explicit new-physics models that introduce such aTGCs, see \citere{Gounaris:2000tb} for instance. 
Indeed, any new fermionic particle can generate $h_3^V$ at one-loop through a triangle diagram, while $h_4^V$ arises
only at a higher loop level or from non perturbative effects as in certain technicolour models.

Our implementation of aTGCs within the $Z\gamma$ \minnlo{} generator follows closely the one in \citere{Campbell:2017aul}. 
The relevant diagrams all involve $q\bar{q}$-initiated topologies where an (off-shell) $Z$-boson or photon is produced in the
$s$-channel and splits into a $Z$ boson and a photon through the anomalous vertices
in \fig{fig:anomvertex}, with a subsequent decay of the $Z$ boson (into charged leptons or neutrinos). 
We stress again that using \eqn{eq:effvertex} and considering the (off-shell) decay of the $Z$ boson  assumes that 
the experiments mostly measure $Z$ bosons close to their mass shell, which is indeed a reasonable assumption.
The relevant tree-level and one-loop amplitudes have been taken 
from \noun{MCFM}~\cite{Campbell:2019dru}, while we extended the calculation of the 
$q\bar{q}\to \ell^+\ell^-\gamma$ and $q\bar{q}\to \nu\bar{\nu}\gamma$ two-loop helicity 
amplitudes of \citere{Gehrmann:2011ab} with the relevant anomalous contributions 
directly within \Matrix{}~\cite{Cascioli:2011va,Buccioni:2017yxi,Buccioni:2019sur}, using the $q\bar{q}V^*$ form factor \cite{Gehrmann:2010ue, Baikov:2009bg, Lee:2010cga} for the loop 
corrections to the tree-level amplitudes with aTGCs.

We now turn to presenting phenomenological results for $pp\to \nu\bar\nu\gamma$ production 
at the LHC with $\sqrt{s}=13$\,TeV centre-of-mass energy for $\nu\in\{\nu_e,\nu_\mu,\nu_\tau\}$. All results have been obtained
with $N_f=5$ massless quark flavours and the corresponding NNLO set of the NNPDF3.0~\cite{Ball:2014uwa} 
parton distribution functions (PDFs) with a strong coupling $\as(\mz)=0.118$. 
The electroweak parameters are evaluated in the $G_\mu$ scheme with the electroweak coupling 
$ \alpha_{G_\mu}=\sqrt{2}G_\mu \mw^2 \sin^2\thW/\pi$ and the mixing angle $\cos^2\thW=\mw^2/\mz^2$.
The input parameters are set to $G_\mu = 1.16637 \times 10^{-5}$\,GeV$^{-2}$, $\mz = 91.1876$\,GeV, $\mw = 80.385$\,GeV,  
$\GZ = 2.4952$\,GeV, and $\GW = 2.085$\,GeV.
The scale setting for \minnlo{} (\minlo{}) is fixed by the method itself and it is the same as in \citere{Lombardi:2020wju}.
In the fixed-order results we set the central renormalization and factorization scales to the transverse mass of the $Z\gamma$ system.
In all cases we use $7$-point scale variations to estimate the uncertainties related to missing higher-order contributions.

\renewcommand*{\arraystretch}{1.2}
\begin{table}[t]
\begin {center}
\begin{tabular}{| c || c | c | c |}
\hline 
  & \setupone{} & \setuptwo{} & \setupthree{} \\
\hline
\hline 
\multirow{ 4}{*}{Photon cuts} &  &  & $\ptg>150$\,GeV \\
& $\ptg>100$\,GeV & $\ptg>150$\,GeV & $|\etag|<1.37$ or\\
&$|\etag|<2.37$&$|\etag|<2.37$&$1.52<|\etag|<2.37$\\
&&&$\dphigmiss>0.4$\\
\hline 
\multirow{ 2}{*}{Neutrino cuts} & \multirow{ 2}{*}{$\ptmiss>90$\,GeV} & \multirow{ 2}{*}{$\ptmiss>150$\,GeV} & \multirow{ 2}{*}{$\ptmiss>200$\,GeV}\\
&&& \\
\hline 
\multirow{ 2}{*}{Jet cuts}  & \multirow{ 2}{*}{---}& Inclusive: $\Nj\ge0$ & \multirow{ 2}{*}{$\Nj\le1$}\\
& & Exclusive: $\Nj= 0$ & \\
\hline
\multirow{ 4}{*}{Jet definition} & anti-$k_{\text{\scalefont{0.77}T}}$ with $R=0.4$ & anti-$k_{\text{\scalefont{0.77}T}}$ with $R=0.4$ & anti-$k_{\text{\scalefont{0.77}T}}$ with $R=0.4$ \\
& $\ptj>30$\,GeV& $\ptj>50$\,GeV &$\ptj>30$\,GeV\\
& $|\etaj|<4.4$  & $|\etaj|<4.5$ & $|\etaj|<4.5$\\
& $\drgj>0.3$ & $\drgj>0.3$  & $\dphijmiss>0.4$ \\
\hline
\multirow{ 4}{*}{Photon Isolation} & Frixione isolation & Frixione isolation & Frixione isolation \\
& $n=1$ &$n=1$ &$n=1$ \\
& $\epsilon_{\text{\scalefont{0.77}$\gamma$}}
=0.5$ & $\epsilon_{\text{\scalefont{0.77}$\gamma$}}=0.1$ & $E_{\text{\scalefont{0.77}T}}^{\mathrm{ref}}=2.45$\,GeV$+0.022\,\ptg$\\
& $\delta_{\text{\scalefont{0.77}$0$}}=0.4$&$\delta_{\text{\scalefont{0.77}$0$}}=0.1$&$\delta_{\text{\scalefont{0.77}$0$}}=0.4$\\
\hline 
\end{tabular}
\end{center}
\caption{\label{tab:setup} Definition of fiducial cuts of two ATLAS measurements, \setupone{}~\cite{Aad:2013izg} and \setuptwo{}~\cite{ATLAS:2018nci}, and of the \setupthree{} that is inspired by the dark-matter search 
of \citere{ATLAS:2020uiq}.}
\end{table}

We consider three sets of fiducial cuts in this letter, which are summarized in \tab{tab:setup}.
The first one (\setupone{}) corresponds to an earlier ATLAS analysis \cite{Aad:2013izg} and 
is used for validation purposes.  To study the effects of aTGCs we use \setuptwo{}, 
which is also employed to compare \minnlo{} predictions to a recent $\nu\bar\nu\gamma$ measurement 
by ATLAS \cite{ATLAS:2018nci}. The last setup (\setupthree{}) has instead been chosen to study 
the importance of NNLO+PS predictions for reducing the uncertainties of the 
$\nu\bar\nu\gamma$ background in dark-matter searches
in the photon plus missing energy channel and it is inspired by a recent 
dark-matter search \cite{ATLAS:2020uiq}.
All three setups include standard cuts on the identified photon and the missing 
transverse energy, a jet definition and Frixione smooth-cone isolation \cite{Frixione:1998jh} for the photon (see \citere{Lombardi:2020wju} for our notation).
In \setuptwo{} one category inclusive over QCD radiation and one with a jet veto is considered.
The \setupthree{}, on the other hand, considers a quite special choice for the smooth-cone parameters
as it combines a fixed (lower) threshold with a fraction of the photon transverse momentum, 
c.f.\ Eqs.\,(2.8) and (2.9) of \citere{Lombardi:2020wju}. Note also that in the \setupthree{} various categories
in $\ptmiss{}$ are considered, as discussed below, 
where the $\ptmiss{}$ cut given in \tab{tab:setup} is just the loosest one.

\begin{figure}[t!]
\begin{center}
\begin{tabular}{ccc}
\hspace{-.5cm}
\includegraphics[width=.34\textwidth]{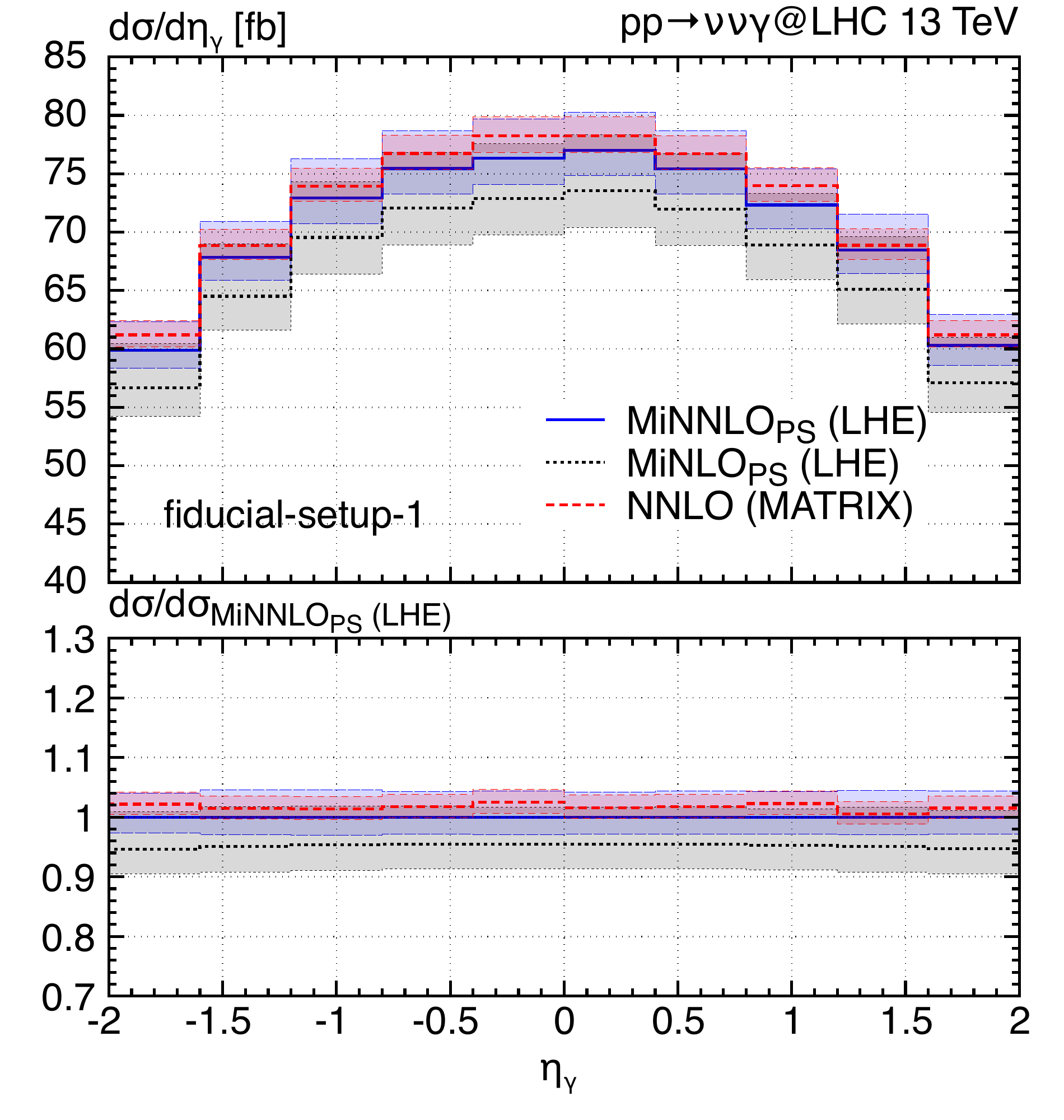}
&
\hspace{-0.55cm}
\includegraphics[width=.34\textwidth]{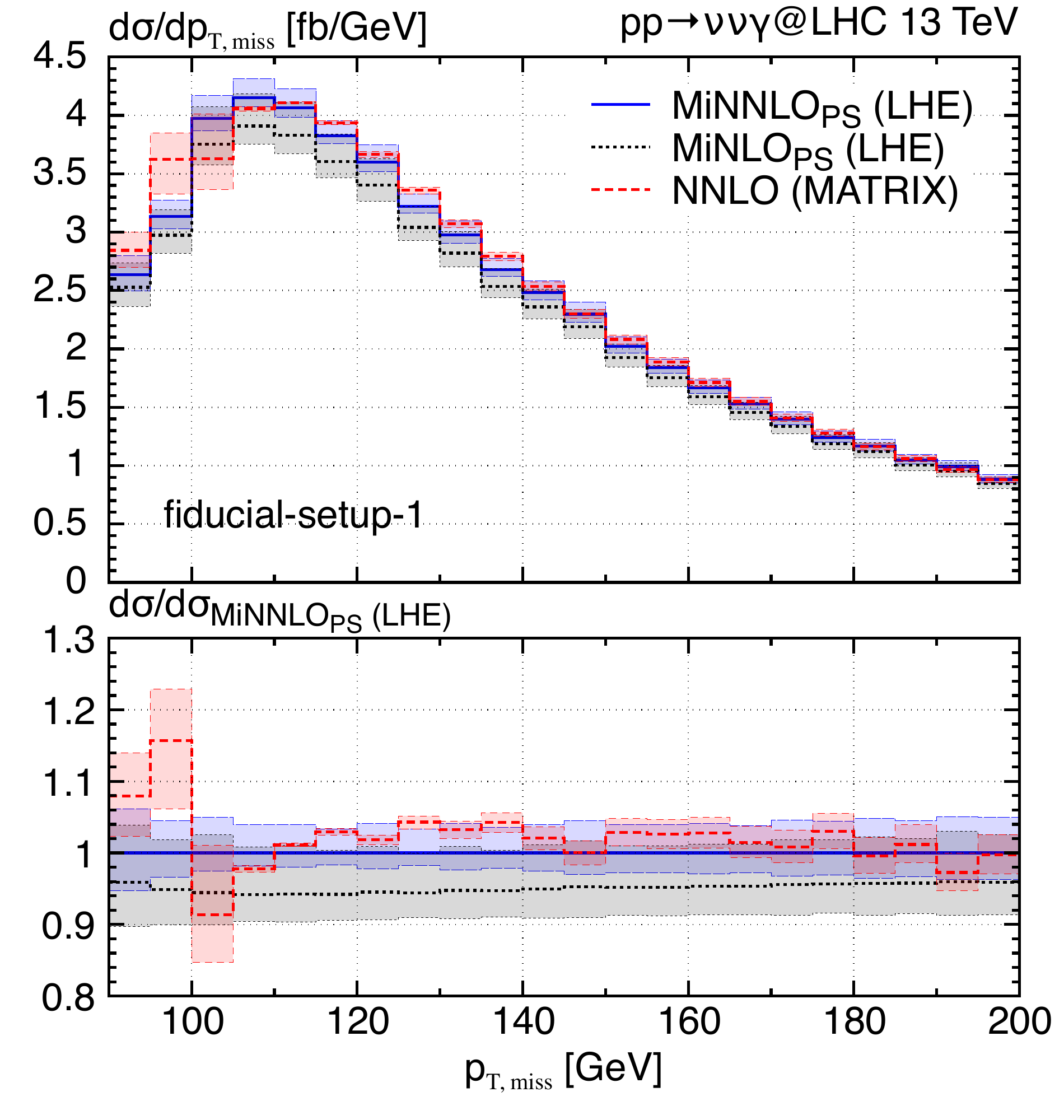} 
&
\hspace{-0.55cm}
\includegraphics[width=.34\textwidth]{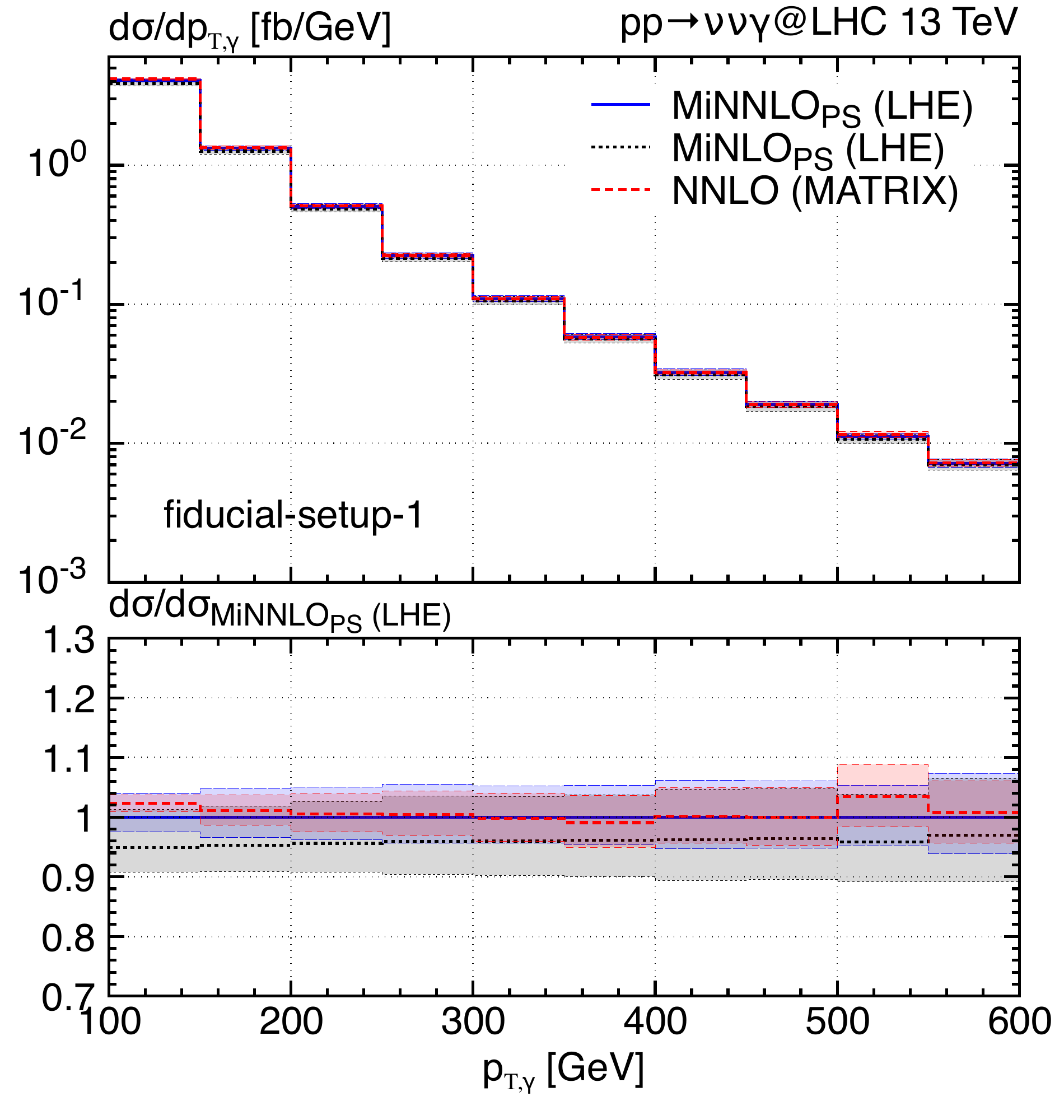}
\end{tabular}
\vspace*{1ex}
\caption{\label{fig:validation} Selected plots at LHE level for validation of \minnlo{} against NNLO.}
\end{center}
\end{figure}

In \fig{fig:validation} we start by comparing \minnlo{} (blue, solid line), \minlo{} (black, dotted line),
and fixed-order NNLO predictions (red, dashed line). 
This comparison is done at the Les-Houches-Event (LHE) level and it serves the purpose of 
numerically validating
the NNLO accuracy of the \minnlo{} predictions as well as indicating the importance of NNLO 
corrections and matching to the parton shower.
The left plot shows the rapidity distribution of the photon. The agreement between \minnlo{} and 
NNLO predictions is excellent, with fully overlapping uncertainty bands.
We remind the reader that small differences (within uncertainties)
between \minnlo{} and fixed-order results are expected due to the different 
scale settings and treatment of terms beyond accuracy, see \citere{Lombardi:2020wju} for instance.
Moreover, when performing scale variations for \minlo{} and \minnlo{}, an additional scale 
dependence is kept in the Sudakov form factor \cite{Monni:2019whf} for a more 
conservative uncertainty estimate, which is absent in fixed-order calculations. This is 
reflected in the slightly larger \minnlo{} uncertainty band. Compared to \minlo{}, however, we observe 
a clear reduction of scale uncertainties from about $6$\% to $3$\% for \minnlo{} and 
roughly an effect of $+3$\% in normalization from the inclusion of NNLO corrections through \minnlo{}.
Also, at high values of the missing energy ($\ptmiss{}$) and of the transverse momentum of the photon ($\ptg{}$) 
\fig{fig:validation} shows that \minnlo{} and fixed-order NNLO predictions are in excellent
agreement. At small values of the missing transverse energy, on the other hand, we observe that 
the NNLO curve develops an interesting feature. This is a 
consequence of the fiducial cut of $\ptg > 100$\,GeV, which induces a perturbative instability \cite{Catani:1997xc} in $\ptmiss{}$
at the threshold, as the region $\ptmiss \le 100$\,GeV becomes sensitive to soft-gluon effects
and is effectively filled only starting from NLO. This behaviour of fixed-order predictions is unphysical
and cured in \minnlo{} (already at LHE level).


We continue by studying the effects of aTGCs on differential distributions.
The search for aTGCs in $Z\gamma$ production has received great attention in the past, 
both at the Tevatron \cite{CDF:2011rqn, D0:2011tjg}, and at the LHC at $7$\,TeV and $8$\,TeV
\cite{Chatrchyan:2011rr,%
	Aad:2012mr,Chatrchyan:2013nda,Aad:2013izg,Chatrchyan:2013fya,%
	Khachatryan:2015kea,Aad:2016sau,Khachatryan:2016yro}.
We stress that, the $Z\to\bar{\nu}\nu$ decay channel has a higher sensitivity to aTGCs due to its higher branching ratio
than the $Z\to\ell^+\ell^-$ decay channel. Indeed,
the most recent $13$\,TeV ATLAS analysis~\cite{ATLAS:2018nci} uses the $\nu\bar\nu\gamma$ final state
to set the most
stringent limits on the aTGCs under consideration thus far,
  which are of the order of $\pm 10^{-4}$ for $h_3^V$ and $\pm 10^{-7}$ for $h_4^V$ (see Table\,8 of ~\cite{ATLAS:2018nci}
  for the exact bounds).
This ATLAS analysis did not make use of any form-factor suppression, which is sometimes applied 
to prevent unitarity violation at high energy, caused by the introduction of aTGCs
at the amplitude level \cite{Green:2016trm}. 
Indeed, in an EFT perspective the terms entering the vertex function in \eqn{eq:effvertex} 
would arise from a set of gauge-invariant operators at dimension-eight (or higher) \cite{Degrande:2013kka}, 
whose validity range is limited by the given new-physics scale. Here, we also 
refrain from using any form-factor suppression.

\begin{figure}[t!]
\begin{center}
\begin{tabular}{cc}
\hspace{-.5cm}
\includegraphics[width=.34\textwidth]{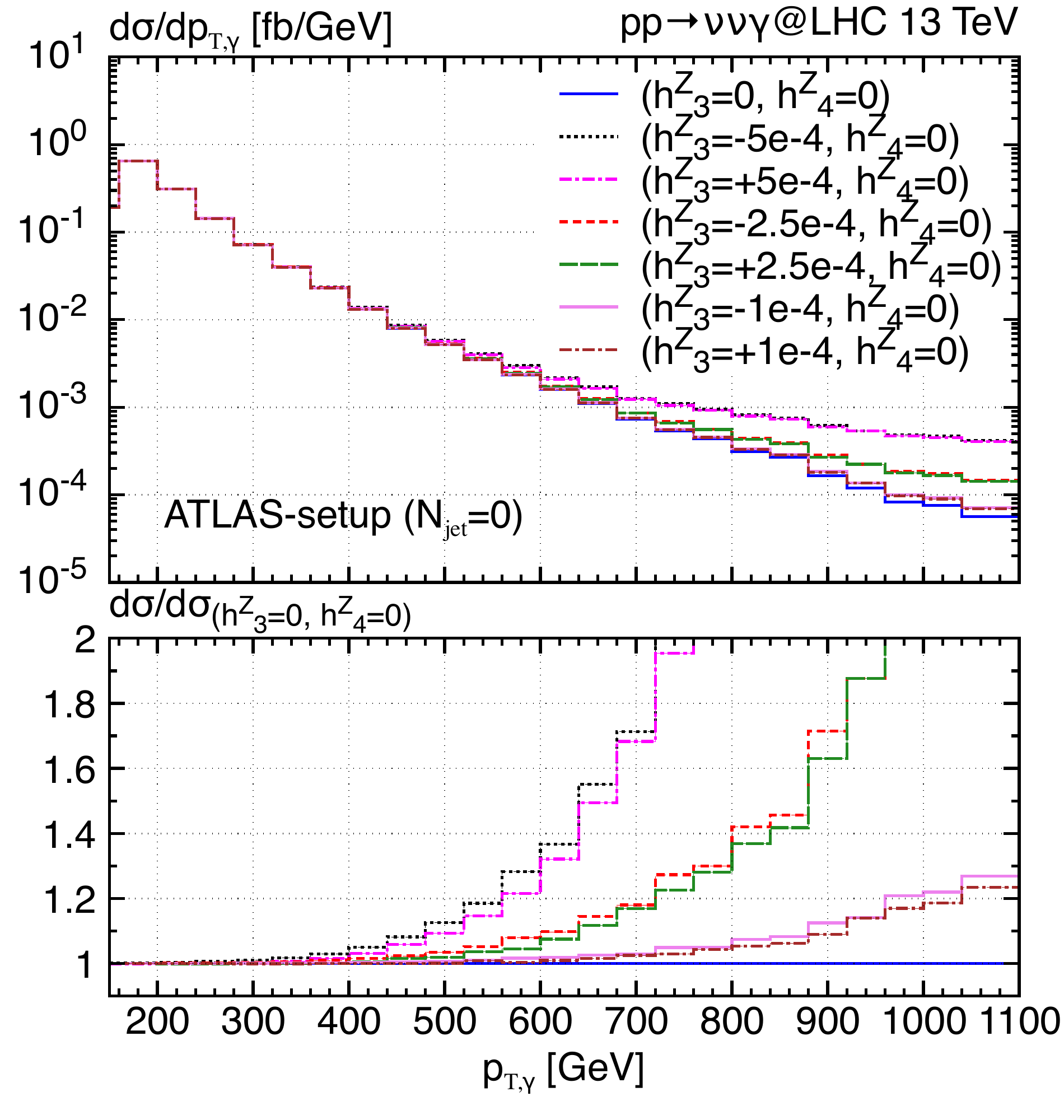}
&
\hspace{+0.75cm}
\includegraphics[width=.34\textwidth]{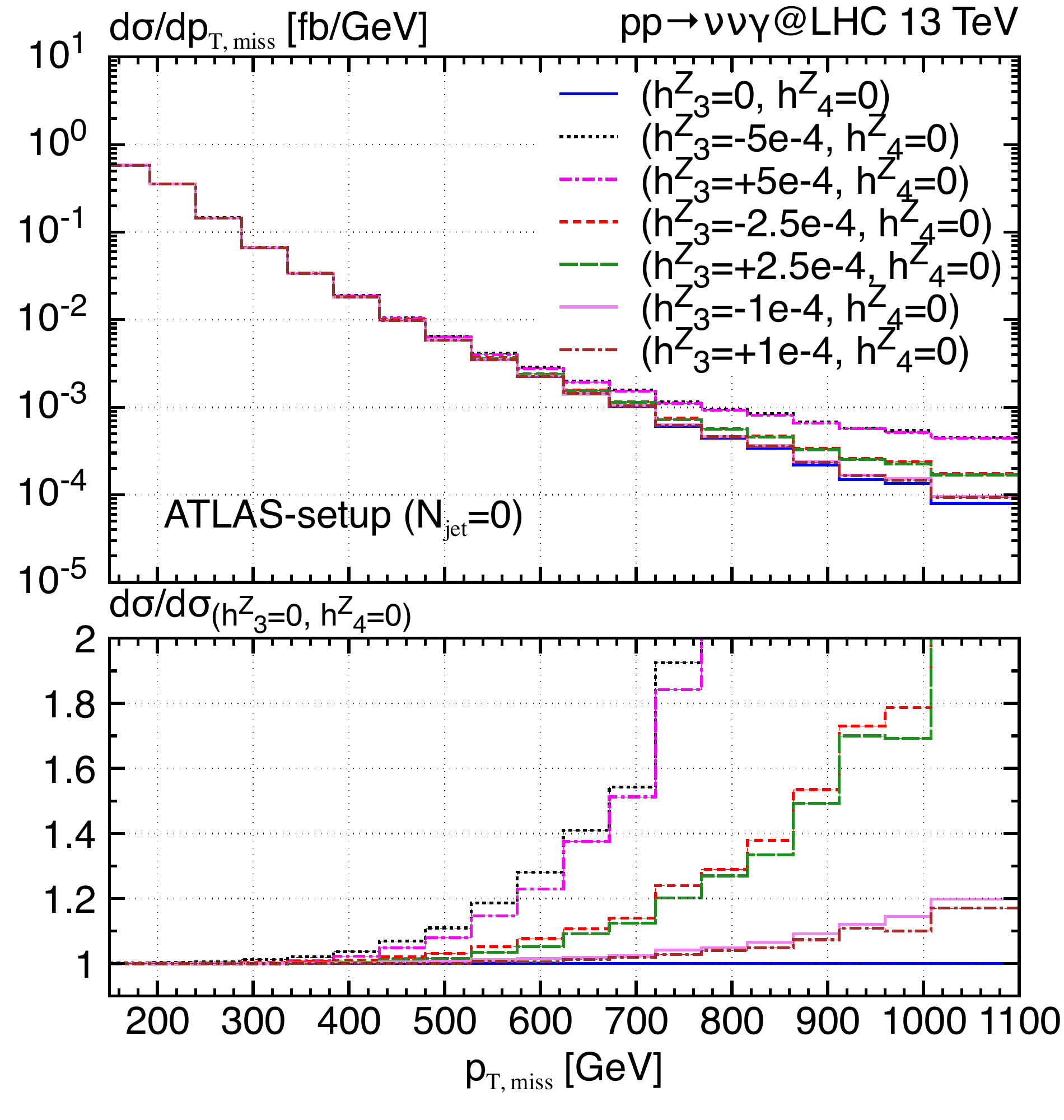} 
\end{tabular}
\begin{tabular}{cc}
\hspace{-.5cm}
\includegraphics[width=.34\textwidth]{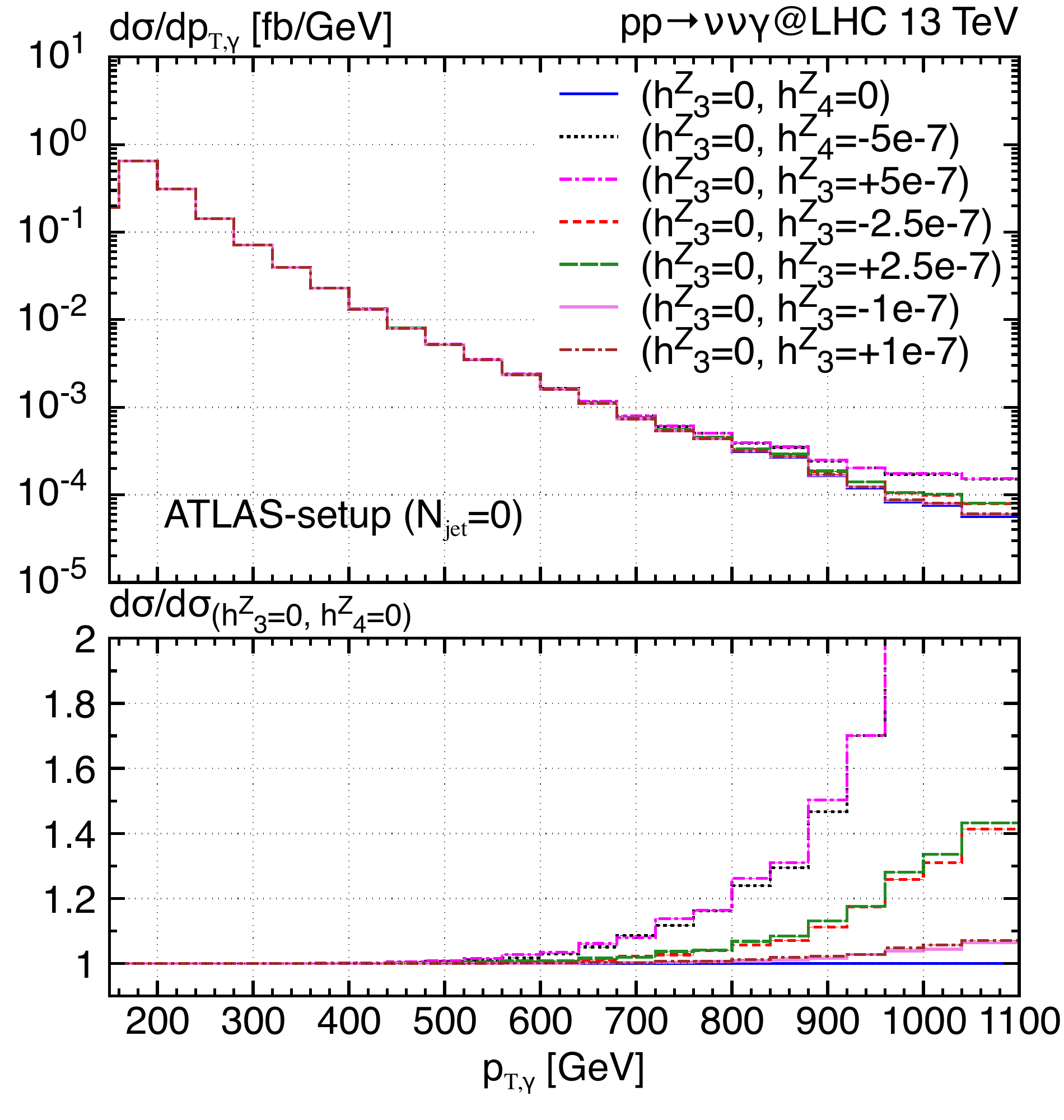}
&
\hspace{+0.75cm}
\includegraphics[width=.34\textwidth]{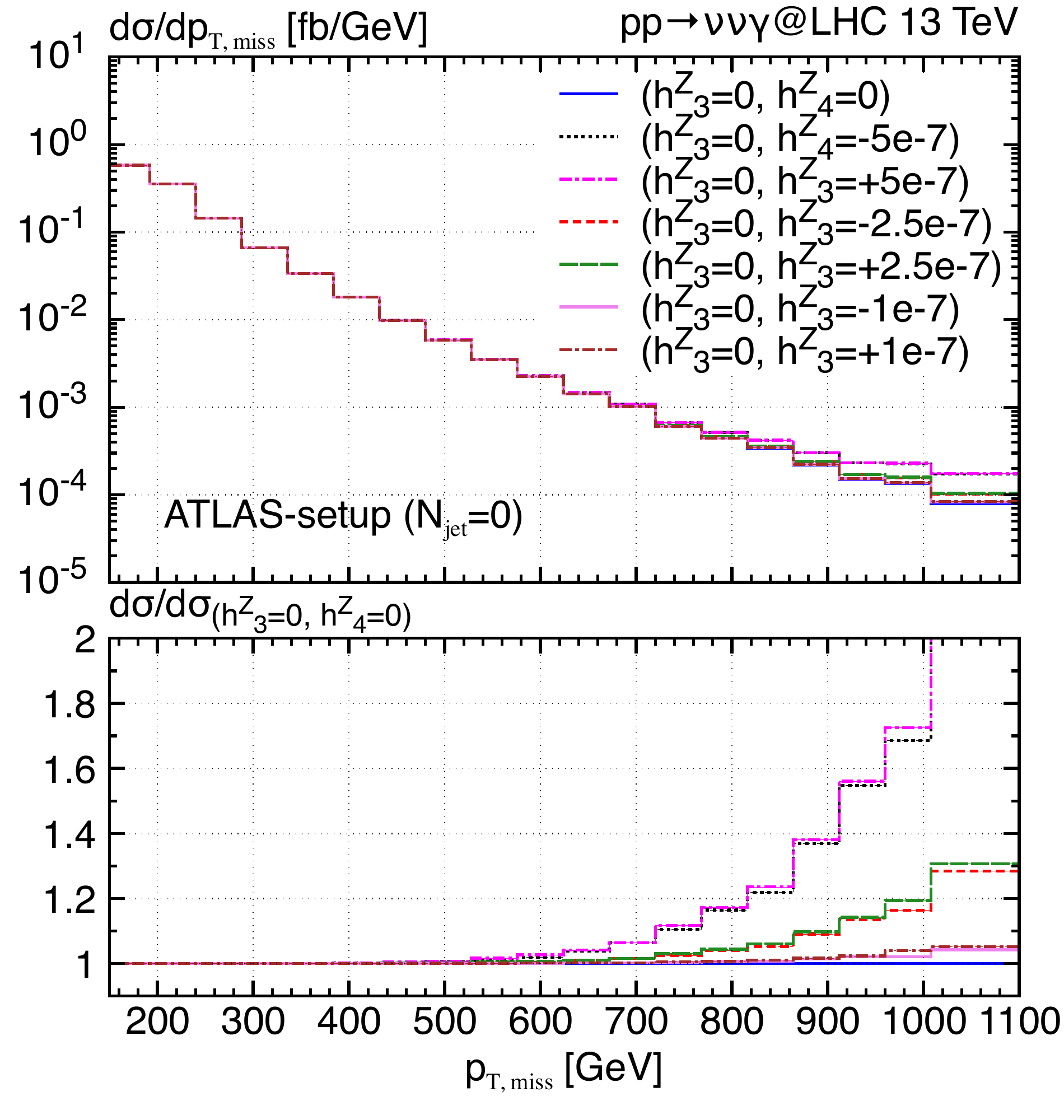} 
\end{tabular}
\begin{tabular}{cc}
\hspace{-.5cm}
\includegraphics[width=.34\textwidth]{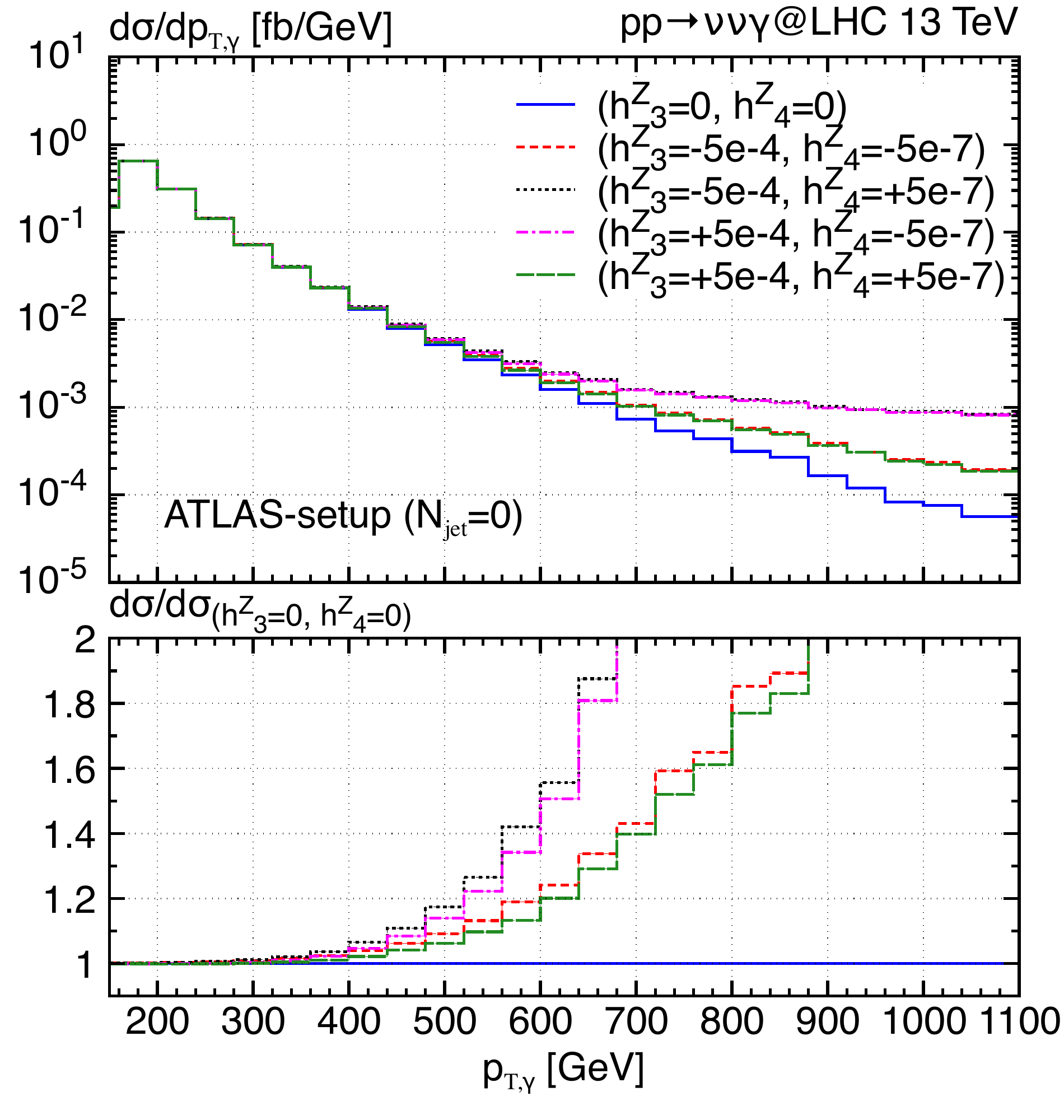}
&
\hspace{+0.75cm}
\includegraphics[width=.34\textwidth]{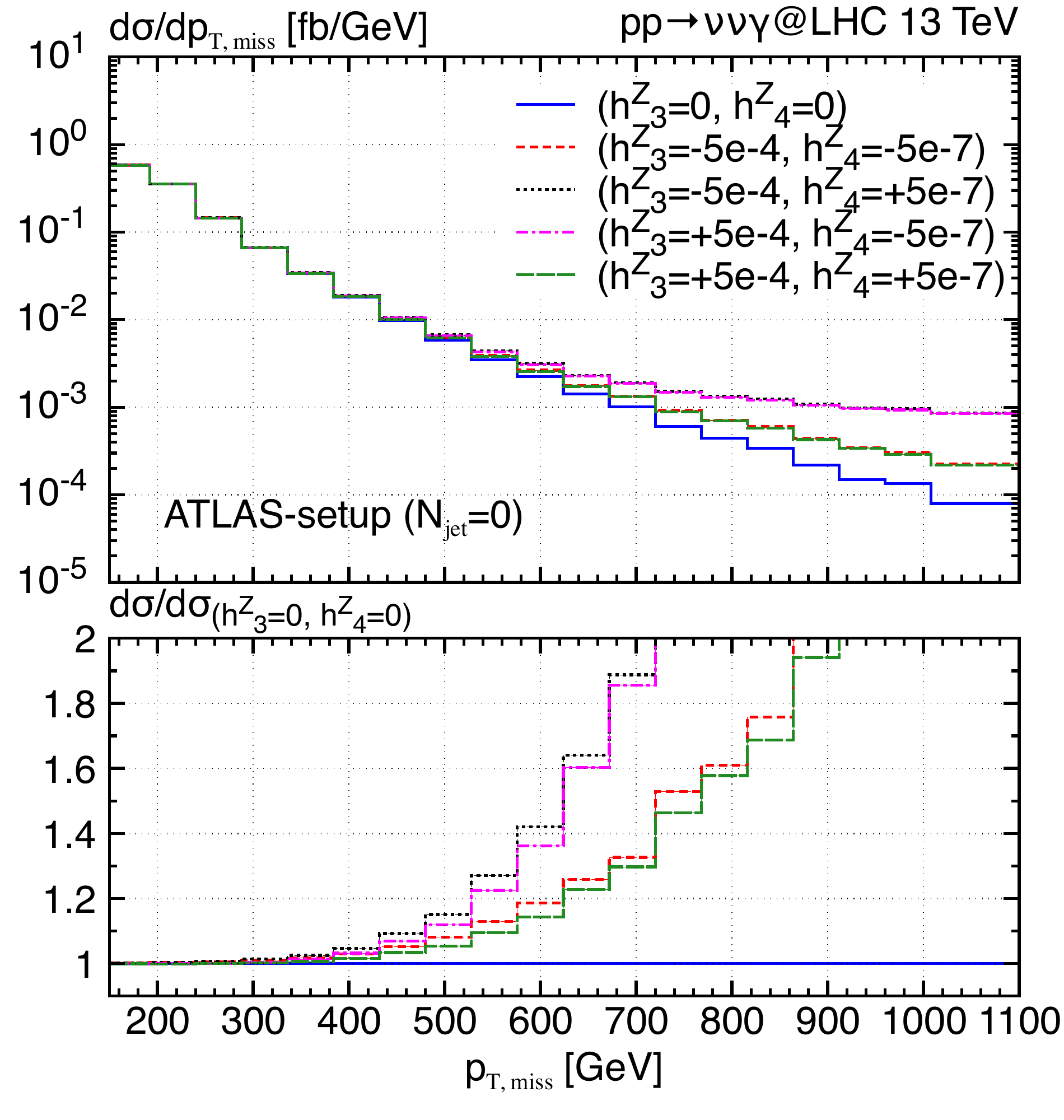} 
\end{tabular}
\vspace*{1ex}
\caption{\label{fig:atgc} \minnlo{} predictions for  the photon transverse momentum (left column) and the missing transverse momentum
  (right column) for different values of $h_3^Z$ and $h_4^Z$. The SM results are reported with a blue, solid line. Individual variations $h_3^Z$
  (two upper plots), individual variation of $h_4^Z$ (two central plots) and their combined variation (two bottom plots) are considered for different values
  within the experimentally allowed ranges defined in Table $8$ of ~\cite{ATLAS:2018nci}. All \minnlo{} results include
parton shower and hadronization effects, as provided by \PYTHIA{8}.}
\end{center}
\end{figure}

Different combinations of the aTGCs have been obtained through reweighting 
at event-generation level (i.e stage $4$ in \POWHEG{}), while ensuring sufficient 
statistics in the relevant phase space regions by accounting for the resonance 
structure of both the $t$-channel SM and $s$-channel BSM 
topologies and by applying a suitable suppression factor to increase the sampling in the high energy tails.\footnote{To this end, we have added the aTGC coefficients as inputs to the \POWHEG{} 
reweighting information, introduced the flag {\tt anommode\,1} that enables the $s$-channel resonance histories associated with 
the aTGCs to be included through the {\tt build\char`_resonance\char`_histories} routines of \POWHEGBOXRES{}, and implemented a suitable suppression factor in the code that can be activated via {\tt suppmodel\,2}.}
Even though all eight anomalous couplings ($h_i^V$ with $i\in\{1...4\}$ and $V\in\{Z,\gamma\}$) 
are consistently implemented in our code, we limit our study to the CP conserving ones,
which do not interfere with the CP violating ones.
Moreover, since also the $V=Z$ and $V=\gamma$ couplings
have been shown to only mildly interfere with each other 
\cite{Baur:1992cd} and to have qualitatively a very similar impact, we focus on the pair $(h_3^Z, h_4^Z)$ here. 
The most relevant phase-space regions to constrain aTGCs in $\nu\bar\nu\gamma$ 
production are the high-energy tails of the $\ptmiss$ and $\ptg$ spectra. 
In \fig{fig:atgc} we show \minnlo{} predictions of each of these two observables in \setuptwo{} 
with a jet veto ($N_{\rm jet}=0$), which is experimentally applied to reduce the SM background
in the tails of the distributions. \minnlo{} results include parton shower and hadronization effects as provided by \PYTHIA{8}~\cite{Sjostrand:2014zea}, with the A14 tune~\cite{TheATLAScollaboration:2014rfk}. We present individual variations of $h_3^Z$ (two upper plots), 
individual variations of $h_4^Z$ (two central plots), and combinations of $(h_3^Z, h_4^Z)$ (two bottom plots), all within the currently 
allowed limits \cite{ATLAS:2018nci}. In all plots, the SM results $(h_3^Z=0,h_4^Z=0)$  are
shown with a blue, solid line.
As one sees from the individual variations of $h_3^Z$ and $h_4^Z$, both negative and positive values of the aTGCs lead to 
a similar positive effect on the spectra, which is a consequence of the very small interference of the $s$-channel 
BSM amplitudes with the $t$-channel SM amplitude, so that the quadratic term in the anomalous couplings dominates. 
This is also the reason why the experimental limits on the aTGCs in $Z\gamma$ production are 
almost symmetric. For values of $h_3^Z$ at the edges of experimentally allowed ranges, we start observing
deviations of  $5$-$10\%$ from the SM for transverse momentum values of $400$-$500$\,GeV, with a steep 
increase afterwards, reaching already $100$\% around $700$-$800$\,GeV.
For $h_4^Z$, whose constraints are at least three order of magnitude smaller,
$5$-$10\%$ effects in the tails of transverse momentum distributions manifest themselves 
starting from $600$-$700$\,GeV, with rapidly increasing effects at larger $\pt$ as well.
  Looking at the simultaneous variations of $(h_3^Z, h_4^Z)$, it is clear that different sign
  combinations constructively interfere, with a mild difference between the two possible sign
  combinations. On the other hand, same sign combinations of $(h_3^Z, h_4^Z)$ interfere
destructively.

\renewcommand*{\arraystretch}{1.7}
\begin{table}[t]
\begin {center}
  \begin{tabular}{lllll}
    \hline
\multicolumn{1}{|c||}{category} & \multicolumn{1}{c}{ SRI1}&  \multicolumn{1}{c}{SRI2}& \multicolumn{1}{c}{SRI3}& \multicolumn{1}{c|}{SRI4} \\[-0.25cm]
\multicolumn{1}{|l||}{$\ptmiss$ [GeV] }  & \multicolumn{1}{c}{$ > 200$} & \multicolumn{1}{c}{ $ > 250$ } &  \multicolumn{1}{c}{$>300$} &  \multicolumn{1}{c|}{$>375$}\\
\hline
\hline
\multicolumn{1}{|c||}{\minlo{} [fb]} & \multicolumn{1}{c}{$27.35(18)_{-3.5\%}^{+6.0\%}$}&  \multicolumn{1}{c}{$12.95(11)_{-4.2\%}^{+6.5\%}$}& \multicolumn{1}{c}{$6.65(8)_{-4.7\%}^{+6.8\%}$}& \multicolumn{1}{c|}{$2.77(6)_{-5.8\%}^{+8.0\%}$} \\
\multicolumn{1}{|c||}{\minnlo{} [fb]} & \multicolumn{1}{c}{$29.09(18)_{-1.9\%}^{+2.9\%}$}&  \multicolumn{1}{c}{$13.77(12)_{-2.2\%}^{+3.2\%}$}& \multicolumn{1}{c}{$7.07(8)_{-2.4\%}^{+3.2\%}$}& \multicolumn{1}{c|}{$2.95(6)_{-3.2\%}^{+4.2\%}$} \\
\hline
\end{tabular}
\end {center}
\caption{\label{tab:dm1} Fiducial cross section in various inclusive $\ptmiss$ categories in the \setupthree{}.}
\end{table}

\renewcommand*{\arraystretch}{1.7}
\begin{table}[t]
\begin {center}
  \begin{tabular}{llll}
    \hline
\multicolumn{1}{|c||}{category} & \multicolumn{1}{c}{SRE1}  & \multicolumn{1}{c}{SRE2} & \multicolumn{1}{c|}{SRE3}  \\[-0.25cm]
\multicolumn{1}{|l||}{$\ptmiss$ [GeV] }  &  \multicolumn{1}{c}{200--250} &  \multicolumn{1}{c}{250--300} & \multicolumn{1}{c|}{300--375} \\
\hline
\hline
\multicolumn{1}{|c||}{\minlo{} [fb]} & \multicolumn{1}{c}{$14.4(1.4)_{-3.2\%}^{+5.4\%}$}  & \multicolumn{1}{c}{$6.30(8)_{-3.7\%}^{+6.2\%}$} & \multicolumn{1}{c|}{$3.88(5)_{-3.9\%}^{+5.8\%}$} \\
\multicolumn{1}{|c||}{\minnlo{} [fb]}& \multicolumn{1}{c}{$15.32(15)_{-1.6\%}^{+2.7\%}$}  & \multicolumn{1}{c}{$6.69(7)_{-2.0\%}^{+3.2\%}$} & \multicolumn{1}{c|}{$4.12(5)_{-1.8\%}^{+2.5\%}$} \\
\hline
\end{tabular}
\end {center}
\caption{\label{tab:dm2} Fiducial cross section in various exclusive $\ptmiss$ categories in the \setupthree{}.}
\end{table}

Now we turn to discussing the importance of NNLO+PS predictions for $\nu\bar\nu\gamma$ 
productions in the context of dark-matter searches in the photon plus missing energy ($\gamma+\Etmiss$) channel,
which is one of (if not the) most important signature to detect dark matter at the LHC, see \citeres{ATLAS:2017nga, CMS:2018ffd,ATLAS:2020uiq}.
Other $X+\Etmiss$ signatures (where X is a visible particle) have been extensively studied
  at the LHC in the past years: for a jet (\cite{ATLAS:2017bfj, CMS:2017zts}), for a heavy quark (\cite{ATLAS:2017hoo, CMS:2018gbj}), for a vector boson (\cite{CMS:2017zts, CMS:2017ret, ATLAS:2018nda})
  and for a Higgs boson (\cite{ATLAS:2017pzz, CMS:2017prz}). In \citere{ATLAS:2020uiq}, which is the most recent dark-matter study in the $\gamma+\Etmiss$ channel, the results are interpreted both in terms of simplified dark matter models \cite{Abdallah:2015ter, Buchmueller:2014yoa, Abercrombie:2015wmb} and of  effective field theories of axion-like particles (ALPs) \cite{Brivio:2017ije}.
As one can see from Table\,4 and 5 of \citere{ATLAS:2020uiq} for instance, the dominant SM 
background is the $\nu\bar\nu\gamma$ process, which also dominates the uncertainties
of the expected SM events. Depending on the category in $\ptmiss$ considered in \citere{ATLAS:2020uiq},
which were used to improve the sensitivity of the analysis,
the uncertainties on the expected $\nu\bar\nu\gamma$ events range from 4\% to almost 15\%.
\citere{ATLAS:2020uiq} based its $\nu\bar\nu\gamma$ predictions on a merged calculation 
of $0$-jet and $1$-jet events at NLO+PS within the {\tt Sherpa\,2.2} MC event generator (\cite{Gleisberg:2008ta, Sherpa:2019gpd}).

Here, we consider \minlo{}  predictions, which have the same formal accuracy as the
merged {\tt Sherpa\,2.2} results quoted in \citere{ATLAS:2020uiq}, and study the reduction of scale uncertainties
when including \minnlo{} corrections in each $\ptmiss$ category. Among the categories in $\ptmiss$ considered
in \citere{ATLAS:2020uiq}, four  are of inclusive type (SRI1-SRI4) and three exclusive (SRE1-SRE3). The
different $\ptmiss$ ranges defining each of the seven categories are summarized in
\tab{tab:dm1} and \ref{tab:dm2}. These tables also report \minlo{} and \minnlo{} predictions for
the cross sections in those categories with the respective
scale uncertainties. Also in this case, results have been showered using \PYTHIA{8}, with the
inclusion of hadronization effects.
One should bear in mind that the experimental analysis is performed at 
the level of events measured in the detector, so that an immediate comparison to 
Table\,4 and 5 of \citere{ATLAS:2020uiq} is not possible. However, both the relative \minnlo{} 
correction and the reduction of scale uncertainties from \minlo{} to \minnlo{} give a good indication
of the expected improvements. Moreover, by and large, relative uncertainties at detector-event level and at
the fiducial level can be assumed to be similar, so that even a direct comparison between 
our fiducial \minnlo{} results and \citere{ATLAS:2020uiq} is not meaningless.
We find roughly a $6\%$ correction in the central value for all categories by including
NNLO corrections through \minnlo{}. Moreover, even though uncertainties 
are larger in categories with more stringent cuts, we still observe an overall reduction
in the uncertainty bands by roughly a factor of two.
Comparing our uncertainties to 
the ones of the $\nu\bar\nu\gamma$ backgrounds reported in Table\,4 and 5 of \citere{ATLAS:2020uiq},
we find smaller uncertainties of \minnlo{} already 
in the SRI1 category ($+2.9$\% and $-1.9$\% compared to $\pm 4.4$\%), while
with increasing $\ptmiss{}$ cut the quoted uncertainties on the $\nu\bar\nu\gamma$ events in \citere{ATLAS:2020uiq}
increase significantly, up to $\pm 13$\% in SRI4, while our \minnlo{} uncertainties amount to less than $5$\%, 
bearing in mind that the translation of the uncertainties of the predictions at detector-event level to
those at fiducial level is not immediate, as explained above.
Similarly, in the exclusive $\ptmiss{}$ categories in \tab{tab:dm2} the \minnlo{} uncertainties stay within about $3$\%, while the 
quoted uncertainties in \citere{ATLAS:2020uiq} of the  $\nu\bar\nu\gamma$ predictions range from $6$\% to $9$\%.
In conclusion, \minnlo{} predictions for $\nu\bar\nu\gamma$ production will allow the experiments to substantially improve the 
dominant background uncertainty in dark-matter searches in the photon plus missing energy  channel.


\begin{figure}[t!]
\begin{center}
\begin{tabular}{ccc}
\hspace{-.5cm}
\includegraphics[width=.34\textwidth]{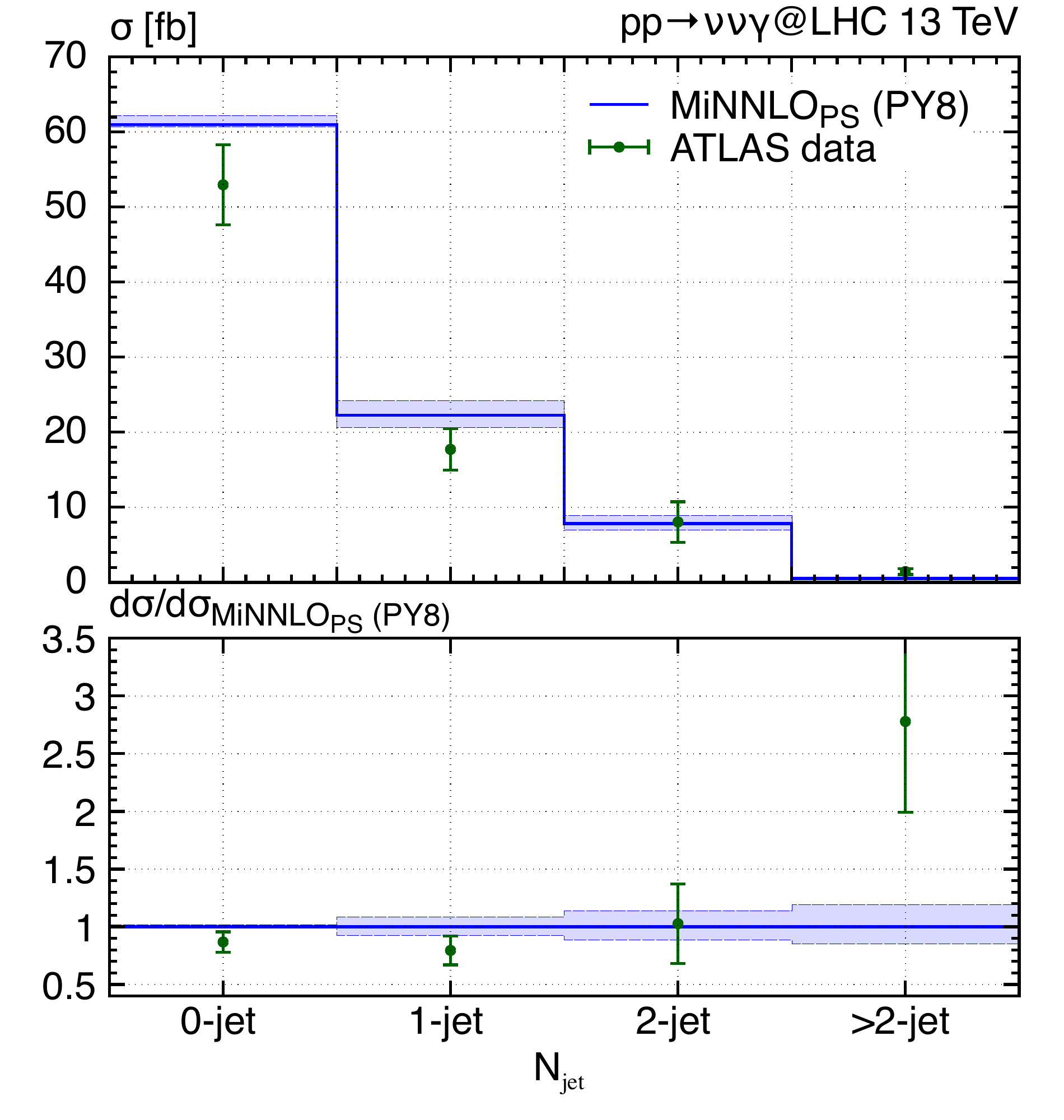}
&
\hspace{-0.55cm}
\includegraphics[width=.34\textwidth]{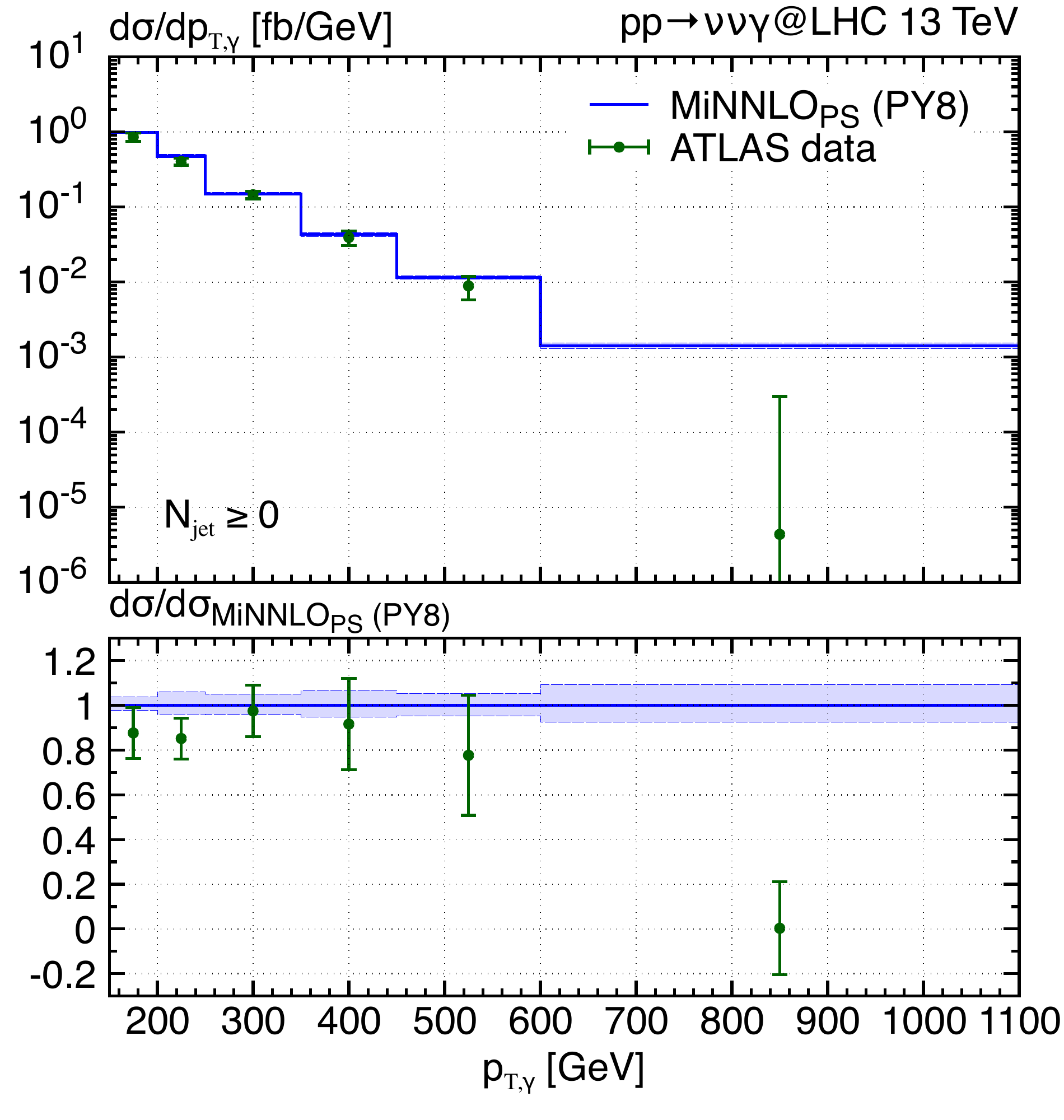} 
&
\hspace{-0.55cm}
\includegraphics[width=.34\textwidth]{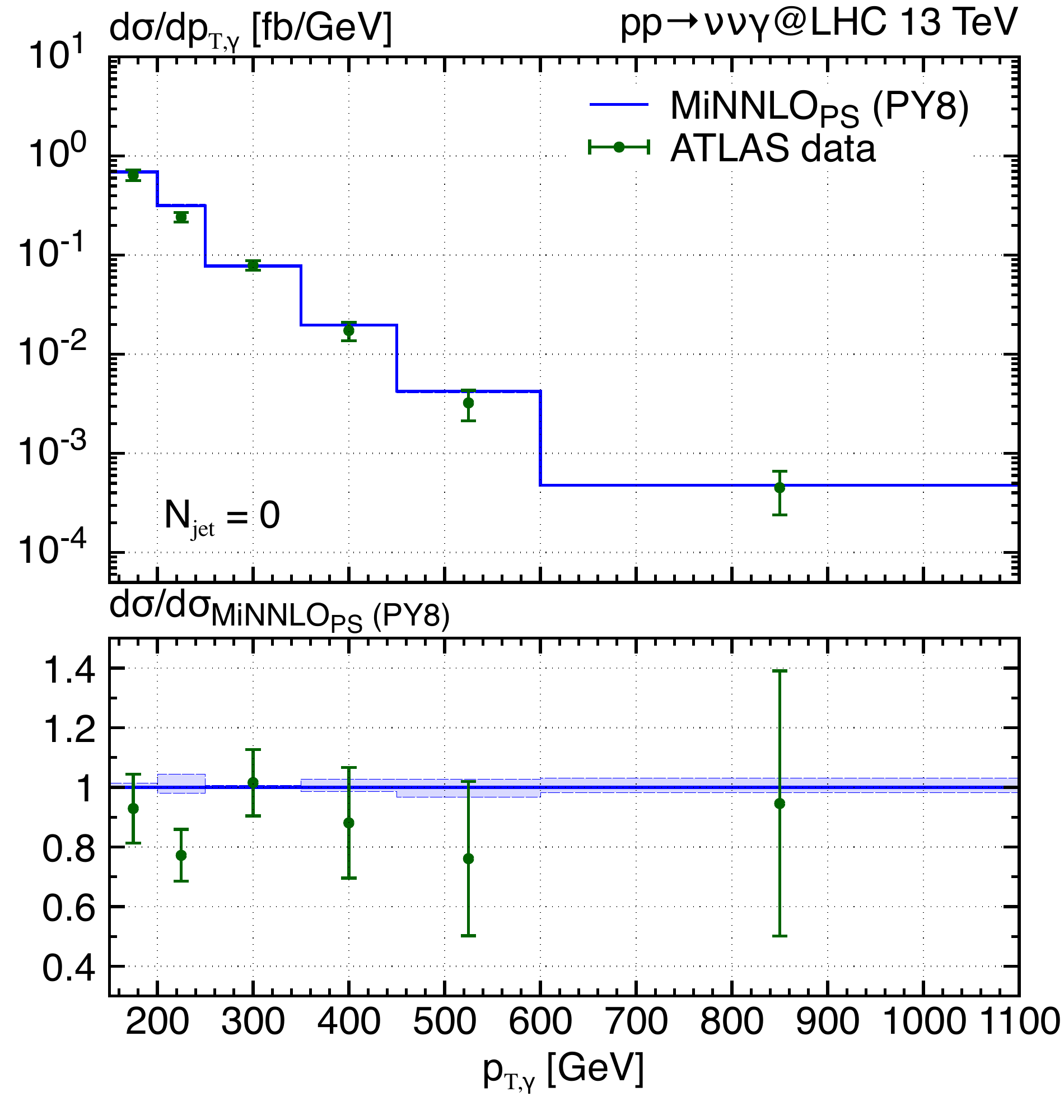}
\end{tabular}
\begin{tabular}{cc}
\includegraphics[width=.34\textwidth]{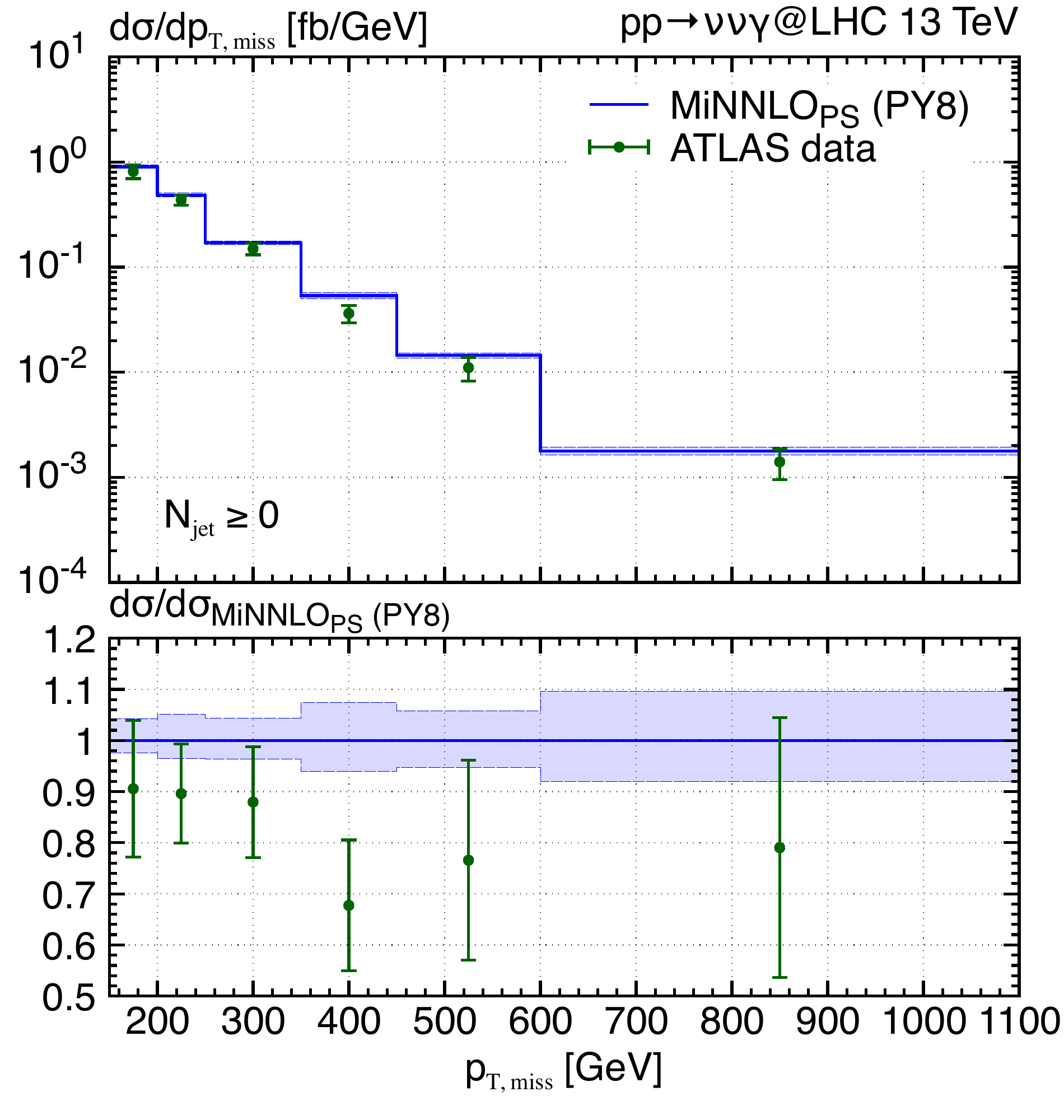}
&
\includegraphics[width=.34\textwidth]{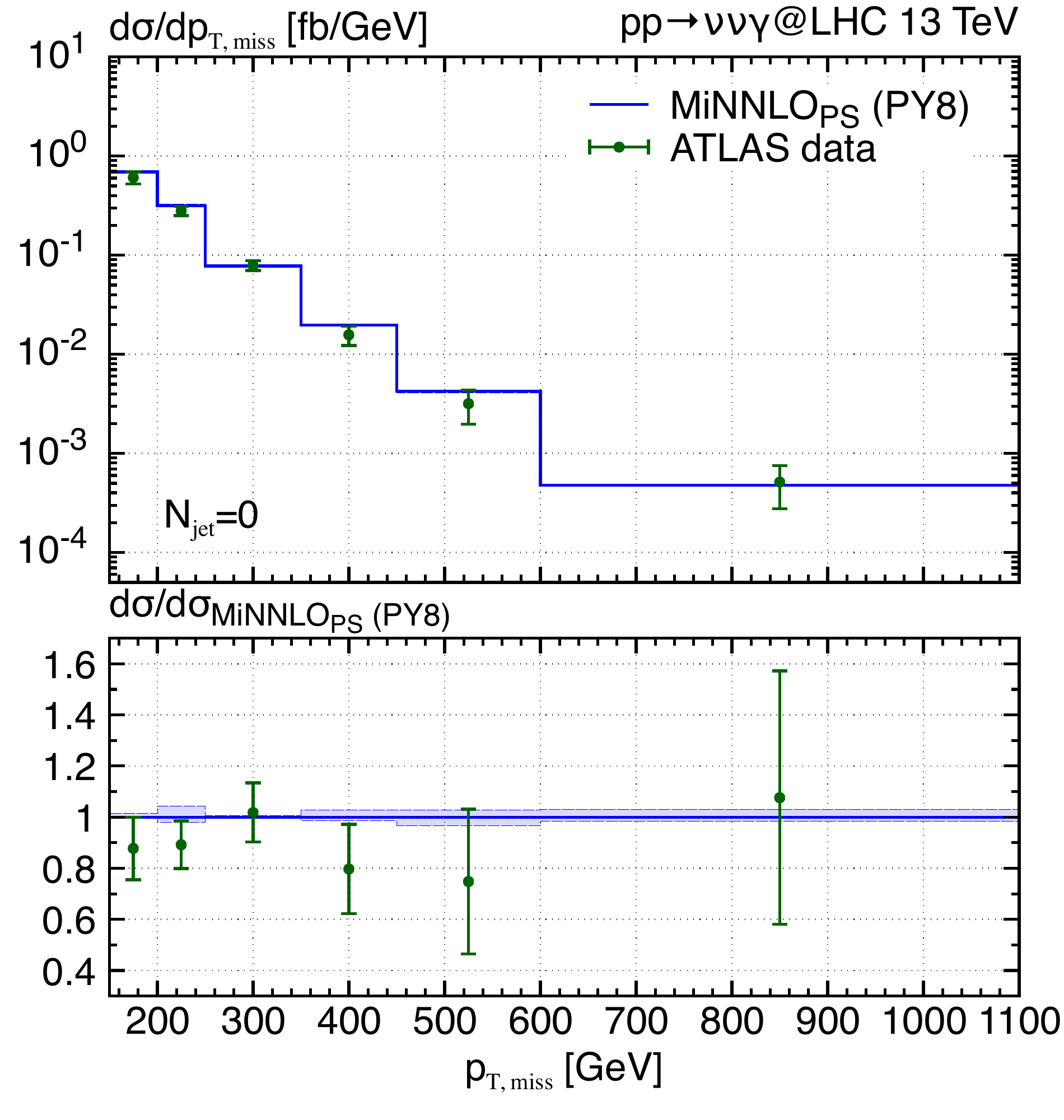}
\end{tabular}
\vspace*{1ex}
\caption{\label{fig:data}  Distribution in the number of jets 
($N_{\rm jet}$), in the transverse-momentum of the photon ($p_{T,\gamma}$)
and in the missing transverse momentum ($p_{T, {\rm miss}}$) compared
to 13\,TeV ATLAS data~\cite{ATLAS:2018nci}. The latter two are shown both inclusively ($N_{\rm jet}\ge0$) and with a jet veto ($N_{\rm jet}=0$).}
\end{center}
\end{figure}

Finally, we compare our $\nu\bar\nu\gamma$ \minnlo{} predictions, including the effects from hadronization,
against ATLAS data \cite{ATLAS:2018nci} in \fig{fig:data}. The first plot shows 
the distribution in the number of jets. The agreement between the \minnlo{}
results and the data points for the different jet cross sections is reasonable, being
within at most two standard deviations. Note that,
starting from NNLO accuracy for the $0$-jet cross section, the accuracy of the 
\minnlo{} calculation decreases by one order for each jet multiplicity, with the 
$>2$-jet multiplicity described only by the shower. This is the reason why  
the prediction undershoots the data in this bin. The other plots in 
\fig{fig:data} show the $\ptg{}$ and $\ptmiss{}$ spectra with and without a jet veto 
($\Nj =0$). Also here the agreement between \minnlo{} predictions and 
data is very good, with deviations of typically one or at most two standard deviations.
There is however one exception: the measured result in 
last bin in the inclusive ($\Nj \ge 0$) 
$\ptg{}$ spectrum is many standard deviations away from 
the prediction. The data point seems to be way too low, when following the 
trend of the distribution. Indeed, it has a very large error and is actually compatible
with zero. Moreover, looking at the $\Nj =0$ result, this bin has actually a higher 
measured cross section than in the inclusive case, which appears inconsistent considering the fact that the $\Nj =0$ cross section should be part of the 
$\Nj \ge 0$ one. A possible explanation could be that, because of the additional jet 
activity, some events are discarded, for instance due to the photon isolation 
requirements. Indeed, looking at the $\ptmiss$ spectrum no such behaviour 
is observed.


To summarize, we have presented the simulation of NNLO-accurate events
for the $Z\gamma$ process including the 
effects of anomalous triple gauge couplings. The calculation 
has been performed in the \minnlo{} framework and we have focused
on the $\nu\bar\nu\gamma$ final state, although the implementation
applies also to the $\ell^+\ell^-\gamma$ process. We followed the 
vertex-function approach for a consistent inclusion of the anomalous couplings, 
but one should bear in mind that there is a direct
translation to an EFT framework \cite{Degrande:2012wf, Degrande:2013kka}.
We validated our \minnlo{} $\nu\bar\nu\gamma$ results numerically 
against fixed-order NNLO predictions, and we have demonstrated
the importance of both NNLO 
accuracy and the matching to the parton shower in certain phase-space regimes.
The effect of the CP-conserving aTGCs has been studied
for the most relevant observables
to extract anomalous couplings in $\nu\bar\nu\gamma$ analyses. 
We found that quadratic terms yield the dominant contribution to the cross section and that
the interference with the SM has a relatively small impact. This typically leads
to very symmetric bounds on the respective coefficients extracted 
by the experiments. We have further shown that the simulation of NNLO-accurate 
$\nu\bar\nu\gamma$ events in the SM is crucial 
to reduce the uncertainty on the dominant background in dark-matter searches 
in the photon plus missing energy channel. Our calculation provides an 
improvement over the
previous scale uncertainties of merged $\nu\bar\nu\gamma$+0,1-jet 
predictions by a factor of two or more.
Finally, the good agreement of \minnlo{} predictions 
with measured distributions underlines the importance 
of NNLO+PS predictions for $\nu\bar\nu\gamma$ production.
We believe that both the inclusion of aTGCs in our $Z\gamma$ \minnlo{} generator 
and the implementation of the $\nu\bar\nu\gamma$ final state
will be very useful for future $Z\gamma$ measurements as well as
BSM searches using the photon plus missing energy signature for aTGCs and dark matter studies at the LHC.

\noindent {\bf Acknowledgements.}
We would like to thank William  Bobadilla, Pier Francesco Monni, Paolo Nason, Emanuele Re, and Vasily Sotnikov for fruitful discussions. 
We have used the Max Planck Computing and Data Facility (MPCDF) in
Garching to carry out all simulations presented here.

\setlength{\bibsep}{3.1pt}
\renewcommand{\em}{}
\bibliographystyle{apsrev4-1}
\bibliography{MiNNLO}

\clearpage

\end{document}

%% file: definitions.tex
\providecommand{\href}[2]{#2}

\newcommand{\tmop}[1]{\ensuremath{\operatorname{#1}}}

\newcommand\F{${\rm F}$}
\newcommand\FJ{${\rm FJ}$}
\newcommand\FJJ{${\rm FJJ}$}

\newcommand{\as}{\alpha_s}

\newcommand{\pt}{{p_{\text{\scalefont{0.77}T}}}}
\newcommand{\GZ}{{\Gamma_Z}}
\newcommand{\GW}{{\Gamma_W}}
\newcommand{\thW}{{\theta_W}}

\newcommand{\Nj}{{N_{\text{\scalefont{0.77}jet}}}}

\newcommand{\ptg}{{p_{\text{\scalefont{0.77}T,$\gamma$}}}}

\newcommand{\ptrad}{{p_{\text{\scalefont{0.77}T,rad}}}}

\newcommand{\ptj}{{p_{\text{\scalefont{0.77}T,$j$}}}}

\newcommand{\Etmiss}{{E_{\text{\scalefont{0.77}T,miss}}}}
\newcommand{\ptmiss}{{p_{\text{\scalefont{0.77}T,miss}}}}
\newcommand{\ptmissvec}{{\vec{p}_{\text{\scalefont{0.77}T,miss}}}}

\newcommand{\mz}{{m_Z}}
\newcommand{\mw}{{m_W}}

\newcommand{\etag}{{\eta_{\text{\scalefont{0.77}$\gamma$}}}}
\newcommand{\etaj}{{\eta_{\text{\scalefont{0.77}j}}}}

\newcommand{\dphigmiss}{{\Delta\phi_{\gamma,\ptmissvec}}}
\newcommand{\dphijmiss}{{\Delta\phi_{j,\ptmissvec}}}

\newcommand{\drgj}{{\Delta R_{\text{\scalefont{0.77}$\gamma j$}}}}

\newcommand{\muF}{{\mu_{\text{\scalefont{0.77}F}}}}
\newcommand{\muR}{{\mu_{\text{\scalefont{0.77}R}}}}

\newcommand{\noun}[1]{{\scshape #1}}

\newcommand{\POWHEG}{\noun{Powheg}}

\newcommand{\POWHEGBOXRES}{\noun{Powheg-Box-Res}}

\newcommand{\minlo}{{\noun{MiNLO$^{\prime}$}}\xspace}
\newcommand{\minnlo}{{\noun{MiNNLO$_{\rm PS}$}}\xspace}

\newcommand{\Matrix}{{\noun{Matrix}}\xspace}
\newcommand{\OpenLoops}{{\noun{OpenLoops}}\xspace}

\newcommand{\PYTHIA}[1]{\noun{Pythia{#1}}\xspace}

\newcommand{\setupone}{{\tt fiducial-setup-1}}
\newcommand{\setuptwo}{{\tt fiducial-setup-2}}
\newcommand{\setupthree}{{\tt DM-setup}}

\newcommand{\citere}[1]{Ref.\,\cite{#1}}

\newcommand{\citeres}[1]{Refs.\,\cite{#1}}

\newcommand{\eqn}[1]{Eq.\,(\ref{#1})}

\newcommand{\fig}[1]{Figure\,\ref{#1}}

\newcommand{\tab}[1]{Table\,\ref{#1}}

\newcommand{\LambdaPWG}{\Lambda_{\rm pwg}}

\usepackage{etoolbox}
\makeatletter
\patchcmd{\@sect}{#8}{\boldmath #8}{}{}
\let\ori@chapter\@chapter
\def\@chapter[#1]#2{\ori@chapter[\boldmath#1]{\boldmath#2}}
\makeatother